# Identification of Neuronal Polarity by Node-Based Machine Learning

Chen-Zhi Su[a,b,1], Kuan-Ting Chou[a,c,1], Hsuan-Pei Huang[a,d],

Chung-Chuan Lo[a,d*], and Daw-Wei Wang[a,b,c,e*]

[a] Brain Research Center, National Tsing Hua University, Hsinchu 30013, Taiwan

[b] Physics Division, National Center for Theoretical Sciences, Hsinchu 30013, Taiwan

[c] Department of Physics, National Tsing Hua University, Hsinchu 30013, Taiwan

[d] Institute of Systems Neuroscience, National Tsing Hua University, Hsinchu 30013, Taiwan

[e] Center for Quantum Technology, National Tsing Hua University, Hsinchu 30013, Taiwan

[1] These two authors contribute equally to this work.

**\* Correspondence: cclo@mx.nthu.edu.tw, dwwang@phys.nthu.edu.tw**

**Abstract**
Identifying the directions of signal flows in neural networks is one of the most important stages for understanding the intricate information dynamics of a living brain. Using a dataset of 213 projection neurons distributed in different regions of a *Drosophila* brain, we develop a powerful machine learning algorithm: node-based polarity identifier of neurons (NPIN). The proposed model is trained by nodal information only and includes both Soma Features *(which contain spatial information from a given node to a soma)* and Local Features *(which contain morphological information of a given node).* After including the spatial correlations between nodal polarities, our NPIN provided extremely high accuracy (>96.0%) for the classification of neuronal polarity, even for complex neurons with more than two dendrite/axon clusters. Finally, we further apply NPIN to classify the neuronal polarity of the blowfly, which has much less neuronal data available. Our results demonstrate that NPIN is a powerful tool to identify the neuronal polarity of insects and to map out the signal flows in the brain's neural networks.

**Keywords** Neuronal Polarity · Machine Learning · Drosophila · Connectome · Axon · Dendrite

**Funding**  This work is supported by the Ministry of Science and Technology grant (MOST 107-2112-M-007-019-MY3) and by the Higher Education Sprout Project funded by the Ministry of Science and Technology and the Ministry of Education in Taiwan.

**Conflict of interests**  The authors declare that they have no conflict of interests.

**Availability of data and material**  The FlyCircuit database **(http://www.flycircuit.tw/)** is provided by the National Center for High-Performance Computing.

**Code availability**  We provide an online version of NPIN to be used or tested by other research groups at the following address:  https://npin-for-drosophila.herokuapp.com/

# Introduction

Rapid technology advances in recent years have led to the development of several connectomic projects and large-scale databases for cellular-level neural images (Chiang et al. 2011; Kuan et al. 2015; Milyaev et al. 2012; Parekh and Ascoli 2013; Peng et al. 2015; Shinomiya et al. 2011; M. Xu et al. 2013). However, how to integrate and transform the data to address scientific questions (Lo and Chiang 2016) remains a central challenge. Overall, these projects aim to provide sufficient information for the analysis of information flows in the brain. This goal is difficult to achieve in the current stage, as many neural images do not provide information on polarity (axons and dendrites). The axon-dendrite polarity of a neuron can be identified by experimental methods (Craig and Banker 1994; Matus et al. 1981; Wang et al. 2004). However, these methods are not practical for large-scale neural image projects and for the image datasets that were already acquired. Morphology-based polarity identification at the post-imaging stage is possible, but this is particularly challenging for insects because of their highly diverse neuronal morphology (Cuntz et al. 2008; Lee et al. 2014).

To address this issue, the method of skeleton-based polarity identification of neurons (SPIN) has been developed using several classic machine-learning (ML) algorithms (Lee et al. 2014). Although SPIN reaches a decent performance in neuronal polarity identification for fruit flies, *Drosophila melanogaster*, with 84%–90% accuracy, the method suffers from the cluster-sorting problem. Most projection neurons (i.e., neurons that innervate more than one neuropil) possess two or more clusters of neural processes. Each cluster can be either axon or dendrite, but not both. Using this observation, the SPIN method first identifies the clusters of processes in a neuron and then identifies the polarity of each cluster. The strategy is highly efficient, but incorrect sorting of clusters can lead to incorrect polarity classification of a large number of terminal points at once. This is a major source of errors in the SPIN method.

In the past decade, modern ML algorithms have been applied in many research fields and in daily life. The popularity of modern ML grows because of rapid developments in computational algorithms, high-speed processors, and big data available from various resources (LeCun et al. 1998; Krizhevsky et al. 2012; LeCun et al. 2015). Some widely successful algorithms —for example, deep neural networks (DNN) and extreme gradient boosting (XGB)— may recognize hidden patterns more efficiently than human knowledge/experience, after proper training on big data. Therefore, ML opens a new era when precise classification and/or prediction becomes possible even without full knowledge of the given data. As a result, many applications of ML have recently appeared in biological and medical research (Asri et al. 2016; Malta et al. 2018; Mohsen et al. 2018). It is reasonable to expect that one may apply modern ML for the identification of neuronal polarity solely using optical images of the fruit fly's brain. For neurons of this insect, several tenths of thousands of high-resolution optical images are already available, which is the largest dataset among all species.

In the present work, we develop a new classifier: node-based polarity identifier of neurons (NPIN). The proposed model achieves much higher accuracy (>96%) than SPIN or the human eye for the identification of neuronal polarity in the *Drosophila* brain. Our NPIN is developed using a node-based feature extraction method. Specifically, NPIN includes both Soma Features (spatial information between a soma and a given node) and Local Features (morphological information around a given node). Two state-of-the-art supervised learning algorithms—XGB and DNN—are used as two complementary classifiers, making the method applicable to complex neurons (which have more than two axon/dendrite clusters) with a competition between Soma Features and Local Features. We further apply NPIN to classify the neuronal polarity of other species of insects (in this case, *Blowfly*), which may have insufficient data for standard ML. Therefore, we find that NPIN provides extremely good results for the classification of neuronal polarity, identifying important local features compared with the known soma features. Moreover, NPIN can be applied to other species. These are all important steps for the understanding of signal flow dynamics in neural networks.

# Method

## Overview

The axon-dendrite polarity of a neuron is correlated with certain aspects of its morphology, such as the distance (or path length) from a terminal to the soma, the number of nodes involved in a domain/cluster, and the thickness of neurites (Craig and Banker 1994; Hanesch et al. 1989; Rolls 2011; Squire et al. 2008). However, so far, very few theoretical frameworks have systematically investigated the relationship between these features and neuronal polarity. These empirical conditions are loosely defined, with many exceptions for different types of neurons. Therefore, it is difficult to identify neuronal polarity by traditional rule-based computational programs. SPIN (Lee et al. 2014), which is developed using classical ML algorithms, can be improved in many aspects.

To overcome these challenges, we significantly improve the previous methods by applying the following four major steps in our NPIN model. It is instructive to briefly describe them (Fig. 1) before the further explanation in the rest of this paper:

**Step I (Data Preparation and Reorganization)**: We invent a diagrammatic method to map a 3D neural skeleton structure of a given neuron onto 2D tree diagrams, called level trees and reduced trees. This effective representation makes it easy to extract representative features for ML.

**Step II (Node-Based Feature Extraction)**: We determine the nodal polarity using the features of each node. Specifically, we identify and extract both Soma Features and Local Features for each node.

**Step III (ML Models)**: In NPIN, we apply two powerful ML algorithms—XGB and DNN—together. They provide two different but complementary approaches for the classification of axons and dendrites.

**Step IV (Implementation of Spatial Correlation)**: The spatial correlation of the nodal polarity in the nearby region is implemented by relabeling the nodal polarity suggested by ML models. This approach can significantly enhance the accuracy of the final output.

Typical ML methods concentrate on the algorithms in Step III. Instead, we put more emphasis on the other three steps in a way specifically useful for the determination of neuronal polarity. Fig. 1 shows the flowchart of the whole calculations. We will explain these strategies in the rest of this section.

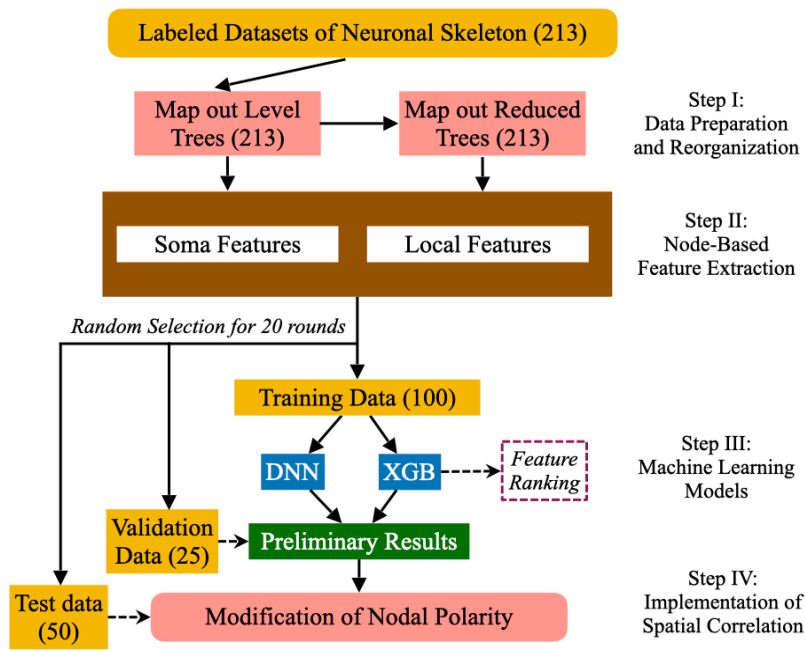

**Fig. 1 Flowchart of the NPIN model.** NPIN includes four major steps, as described in the text. The dataset contains 213 neurons with labeled polarity as the ground truth. We randomly choose 100/25/50 neurons from the datasets for training/validation/test sets. Every neuron in the training/validation sets is mapped to a level tree and a reduced tree. We then extract Soma Features and Local Features from these neuronal data for training. Preliminary results are obtained by XGB and DNN algorithms after validation. We then relabel the classification by including spatial correlations of nodal polarities before comparing them with the test data with known polarities. The whole process is repeated 20 times to cover all 213 neurons in the original dataset. As a result, each neuron could be selected to be a test sample and classified by a model trained on other neurons.

# Dataset

Our main dataset represents 213 neurons in the *Drosophila* brain, which are available from the FlyCircuit database (http://www.flycircuit.tw/) (Chiang et al. 2011). These 213 neurons are all projection neurons selected from various regions across the brain to represent the diversity of neuronal morphology as much as possible (Fig. 1(a)). These neurons innervate 15 neuropils: AL, AOTU, CAL, CCP, DMP, EB, FB, IDFP, LH, LOB, MED, NO, PB, VLP, and VMP. Among these 213 neurons, 107 neurons have been included in the dataset used in the development of the previous model, SPIN, and we have 106 additional neurons for the present work. As we will show later, due to the improvement of feature extraction and the ML algorithm, our model, NPIN, substantially outperforms SPIN, not only in the overall precision and recall but also in the applicability in more brain regions as well as more types of complex structures. In Appendix E, we list these 213 neurons with information including the brain regions innervated by the dendrites and axons of each neuron, the number of axon and dendrite terminals, and precision/recall obtained by our model.

We divide the neurons in our dataset into two types: (i) simple neurons, which have two clusters of terminals (one dendrite and one axon); (ii) complex neurons, which have more than two clusters of terminals. In Figs. 2(b1)–(b3) and (c1)–(c4), we show some typical skeleton structures of these two types of neurons. In our dataset, we have 89 simple neurons and 124 complex neurons with previously reported polarity. Among complex neurons, most complex neurons have three clusters (two dendrites and one axon, or one dendrite and two axons). Only a few neurons have more than three clusters. The reason to classify these neurons is to investigate how the distance to soma and the number of clusters can influence the identification of neuronal polarity. Moreover, we can examine how well NPIN performs even when the polarity is difficult to be identified by the human eyes in the case of three or more terminal clusters. This is one of the most important criteria for a polarity identifier to be practically applicable for the determination of signal flow in neuronal networks of the insect brain. There are, of course, some other types of projection or local neurons, which may not be easily classified by the number of clusters or by their polarity distribution. We do not include them in the dataset of this study because of a lack of data with confirmed polarity to be used for training. Our approach developed here, however, may still be applicable to these neurons when more data are available in the future.

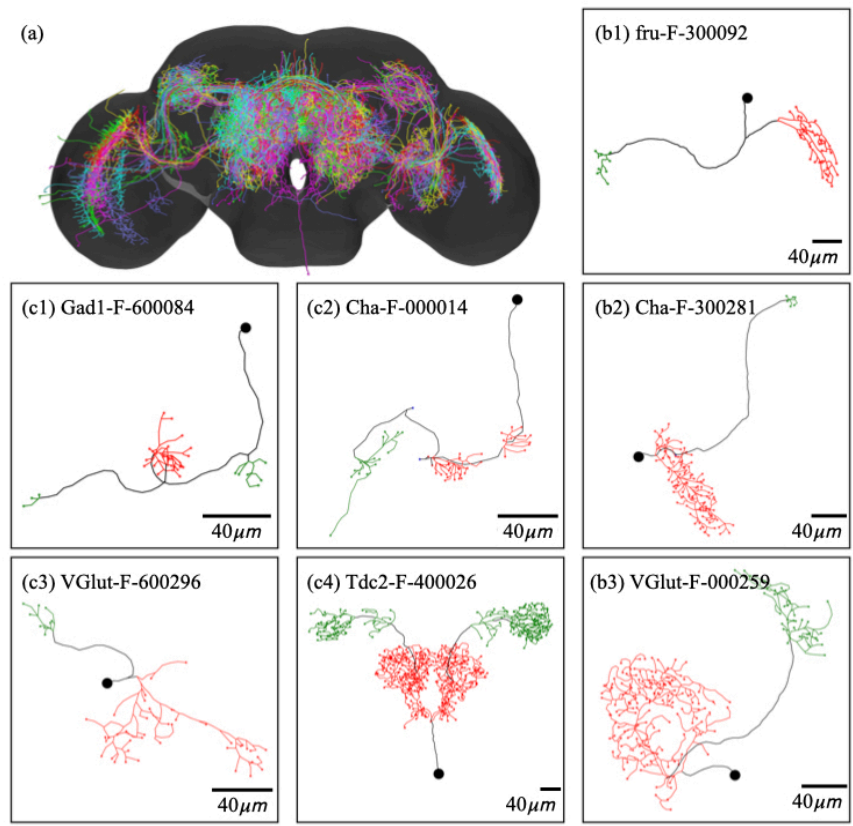

**Fig. 2** ***Drosophila melanogaster* (fruit fly) neurons used in the present study.** (a) All 213 neurons in our dataset, shown in their actual locations in the standard fly brain. (b1)–(b3) Skeleton structures for several simple neurons. (c1)–(c4) Skeleton structures for several complex neurons. Black dots represent somas. Black lines are the main trunks of neurons. Green or red lines indicate the axonal or dendritic clusters, respectively. Each neuron is labeled by its ID in the FlyCircuit database.

## Standardized Representation: Level Trees and Reduced Trees

To improve the accuracy of our ML model, we first need to define how to "standardize" the morphological information of these neurons, which are so different from each other in their original 3D structures. Figs. 3(a1) and (b1) show two examples of a simple neuron and a complex neuron. First, we start with the 3D skeleton structures (see Figs. 3(a2) and (b2)) extracted from the raw images, where the width information of the trunks or branches are ignored. In our work, we further map the 3D skeleton structure onto a level tree (see Figs. 3(a3) and (b3)), which keeps all information on the position of each node (including soma, terminals, and cross points between branches) and the path length between them, but it ignores the trunk and branch information, such as width or shapes. To express this information in a 2D diagram, we introduce the level structure according to the generation of nodes: a soma is placed in the top-level (level 0), and the next two nodes are placed in the lower level (level 1), and so on for their offspring, until all the ending nodes (terminals) are properly placed. We take the convention that the branches with more successive non-empty levels are placed in the left-hand side of the branches with less successive non-empty levels (Figs. 3(a3) and (b3)). We believe that most morphological features of the neuronal cluster are still extractable from such standardized representation because the spatial positions of all nodes (including a soma and terminals) are still available. The only missing information in the level tree (compared with the raw 3D image of neurons) is on shapes and widths of neuronal branches that connect neighboring nodes. As we will see below, this missing information is not important for the determination of neuronal polarity.

In addition to the level tree representation for a neuronal structure, in this work, we further define a reduced tree for each neuron. The reduced tree aims to retain the major branches of the skeleton structure to identify an axon or dendrite cluster. This information is important for the determination of cluster curvature and aspect ratio for nodal features within each cluster (explained below). The reduced tree of a neuron can be obtained by repeatedly removing the ending nodes with the branches shorter than a characteristic length determined by the branch distribution, until it stops automatically or only five levels are left (see Figs. 3(a4) and (b4)). The basic assumption behind this procedure is that the major branch of a neuron skeleton structure is contained in the "inner" (closer to the soma) and "longer" branches. Shorter and outsider branches are minor or unimportant for determining the clusters. See Appendix A for the detailed procedure of producing the reduced tree from a level tree.

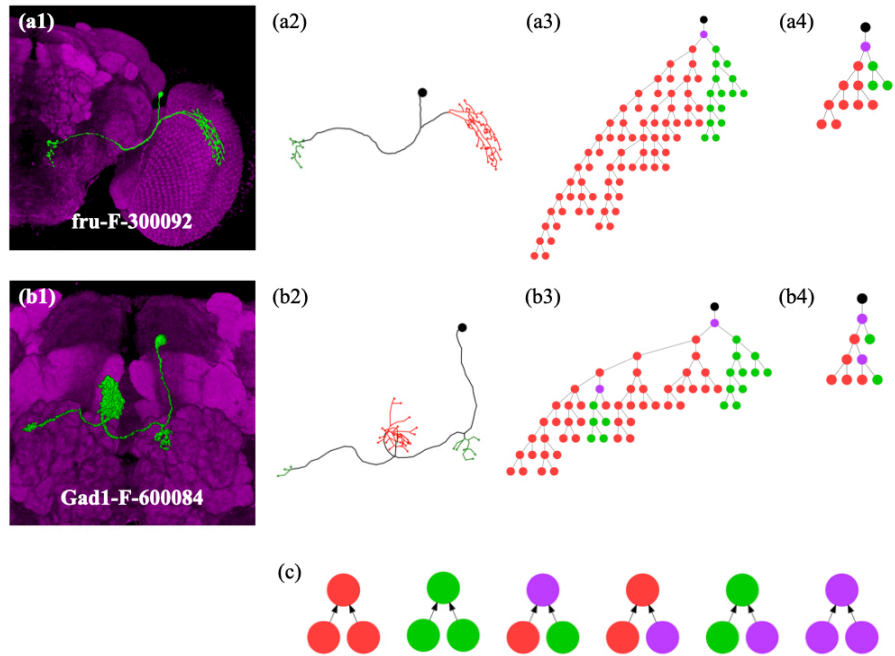

**Fig. 3 Encoding 3D optical images of neurons into level trees and reduced trees.** First, the volume image of a neuron (a1) is converted into the skeleton (a2), and then a level tree (a3), which is a 2D plot with a standardized method to label most features of the original neurons. Red, green, and purple dots represent dendrites, axons, and dividing nodes (including terminals), respectively. (a4) represents the reduced tree of the same neuron cell. (b1)–(b4) show the same reduction for a complex neuron. Because a complex neuron has more than two clusters, there can be more than one dividing node that separates axon clusters from dendrites. In (c), we graphically show the rules to define the nodal polarity based on the polarity of terminals in the level tree (see the text). Upward arrows indicate that the polarity in the upper level is determined by the polarities of the two nodes/terminals in the lower level.

## Nodal Polarity

The polarity of the neurons in our dataset are all predetermined using the presynaptic (Syt::HA) or postsynaptic (Dscam17.1::GFP) markers (C.-Y. Lin et al. 2013) or using the morphological features described in previous studies (Fischbach and Dittrich 1989; Hanesch et al. 1989; Wu et al. 2016). There are 7142 terminals identified as dendrites and 2310 as axons. However, because the axon-dendrite polarity of these terminals is highly correlated to the morphological structure of their neurons, in this study, we extend the definition of polarity from terminals to nodes, and we use this information to extract features in NPIN.

We emphasize that using features extracted from nodes has several important advantages over using features extracted from terminals or clusters for the training process of ML. First, the number of nodes is much larger than the number of clusters. Therefore, the polarity identification has significantly higher accuracy due to the larger training sample. Second, nodes are well-defined in the skeleton structure (compared with clusters) and could include more morphological features (compared with terminals). Finally, these nodes can also be systematically labeled in the skeleton structure or in our level tree diagram, making it easy to include their correlated features in the spatial distribution. This node-based feature extraction is crucial in NPIN, making an accurate identification of neuronal polarity possible.

To extend the polarity definition from terminals, as provided in the dataset, to nodes on the skeleton of a neuron, we apply the following series of rules to define the nodal polarity according to the polarity of terminals (Fig 3. (c)): (1) If two child nodes (or terminals) are both axons (or dendrites), their parent node (the node that directly connects to them in the upper level) is also defined as an axon (or dendrite). (2) If one of the child nodes (or terminals) is an axon, and the other is a dendrite, their parent node is defined as a "dividing node." (3) If one of the child nodes is an axon (or dendrite), and the other is a dividing node, their parent node is defined as an axon (or dendrite). Finally, (4) if two child nodes are both dividing nodes, their parent node is also defined as a dividing node (however, we do not have such a case in our dataset).

After applying these rules, we can label the polarity of all nodes of any neuron using the polarity information of their terminals. We do not consider this procedure as introducing artifacts or unconfirmed polarity labeling. The reason is that the polarity is typically a property of an entire cluster (or domain) of branches in a projection neuron, not only its terminals. We note that the dividing node is defined to mark the position to separate axon and dendrite clusters, and it should be important in the nerve cell development. Since the number of dividing points is much less (one or at most two points in each neuron) than the number of axon or dendrite nodes, we do not include them in the training and testing processes. Figs. 3(a3) and (b3) show some representative level trees, where all nodes are properly labeled.

## Feature Extraction for Nodal Polarity

In principle, the level tree representation defined above contains all information of a 3D neuron and can be used for the identification of neuronal polarity. However, due to a large number of features and their diversity, a comprehensive understanding of the prediction/classification will be very difficult. As a result, we will use only certain representative features for ML, so that the obtained result can be understood and interpreted more easily.

In this work, we define two types of representative features for polarity identification: Soma Features (SF) and Local Features (LF). Soma Features contain certain spatial information from a given node to a soma, including the path length along the neuronal branches and the direct distance in 3D space. Local Features contain certain information on the local morphology of a given node, including the curvature and aspect ratio of the cluster it belongs to. Hence, Local Features do not include any information about the soma, while Soma Features do not include any information about the local morphology. Let $i$ be the index of a given node. Soma Features of node $i$ can be expressed as a four-component vector: $SF = [l_{si}, nl_{si}, d_{si}, nd_{si}]$. Local Features of node $i$ can be expressed as a five-component vector: $LF = [l_{pi}, nl_{pi}, c_i, ar_i, rl_i]$. If an offspring node does not exist, its features are replaced by number $-1$. We then train different ML models on various combinations of features to identify their roles in the identification of neuronal polarity. In Appendix B, we explain how to identify and calculate soma features and local features (from the level trees and reduced trees defined above).

### ML Models

We train our model by supervised learning using the training data extracted from the dataset. We implement several ML algorithms: random forest, gradient boosting decision tree, XGB, support vector machines, and DNN. We find that, in general, XGB and DNN provide the best and complementary results from the features we selected. Therefore, we use them in our NPIN. In Appendix C, we explain the details of how to implent these two algorithms in the present study.

In addition to the algorithms, an ML model also depends on the features used during the training process. To investigate the effects of different morphological features on the identification of nodal polarity, we develop three models by using three types of features in NPIN: Model I (using both Soma Features and Local Features), Model II (using Soma Features only), and Model III (using Local Features only). As we will see later, we can gain insight into the relationship between morphological features and polarity by systematically comparing the polarity identification results between different models and different types of neurons.

### Implementation of Spatial Correlation of Nodal Polarity

In the standard application of supervised learning for classification, one usually obtains the results from the output probabilities directly when the model is well-trained on the training data. The training aims to minimize the cross-entropy between the output results and the known answers by backpropagation. However, this ML process does not guarantee reasonable results all the time without violating some necessary conditions, which could not be included in the input features of training data. For the task of nodal polarity identification in our present work, for example, the polarities of nodes are highly dependent on its neighboring nodes: nodes in the same cluster (and, therefore, close in space) are usually of the same type (a dendrite or axon), but such loosely defined necessary condition cannot be implemented in the loss function if the polarity of each node is identified individually. Therefore, we have to include such a spatial correlation of polarity by adding other methods in the ML model.

In this work, spatial correlations between nodal polarities can be included by the modification of the polarity provided by XGB or DNN, if the probability for axon or dendrite is below a certain threshold. More precisely, such a modification process contains three steps: (1) we perform the ML process for the test data and obtain the polarity and its probability for each node. (2) Next, we accept the result of a given node if the probability is higher than a threshold, and we reject the result otherwise by changing it to be unidentified. (3) Finally, we relabel these rejected/unidentified nodes according to the polarity of its neighboring nodes. As a result, we identify spatial correlations between nodal polarities. More details of such polarity modification and its effects on the NPIN performance are described in Appendix D.

# Results

Our dataset includes 213 neurons with verified polarities as the ground truth. In our training procedures (Fig. 1), we randomly select 100 neurons from the dataset for training, 25 for validation, and 50 for testing. This process is repeated for 20 rounds, so that each neuron can be tested (by different models trained by other neurons) for 4–5 times on average. We then average these probabilities for their nodal polarity and make the final comparison with the ground truth. Using this method, the obtained results for the nodal polarity of each neuron can be much more stable because the fluctuations due to the dataset selection are reduced. In our training data and in comparison with the ground truth, the dividing points are not included because their numbers are too few to be statistically relevant. In the testing neurons, they could be recovered using the predicted polarities of other nodes (see Appendix D).

In the following sections, we will first present the distribution of nodal features, including both soma features and local features, obtained from all neurons in our dataset. This provides a deep understanding of neuronal morphology and its relationship with other results. Next, we show the results of polarity identification provided by Model I (with both Soma Features and Local Features) for our whole neuron dataset, followed by results using Model II (with Soma Features only). We then focus on the results obtained by using complex neurons as training data for comparison. As an example of application in other species, we apply NPIN to test the blowfly. Finally, we summarize these calculation results and our findings.

## Feature Distribution and Importance Ranking

Before presenting the results of neuronal polarity by NPIN, we investigate the distribution of different features (Soma Features or Local Features) for different types of neurons (simple neurons or complex neurons). This provides a better picture which helps to understand and explain the results of the present algorithm. In Fig. 4, we show the distribution of axon nodes and dendrite nodes (including terminals) of all neurons as a function of the normalized path length (relative to the largest length to the soma). Results of simple neurons (a1) and complex neurons (a2) are shown together for comparison. As expected, most axons have a longer path length to soma compared with most dendrites in simple neurons, but the distribution of dendrite is certainly wider than the distribution of axons. A wider distribution pattern for dendrites in simple neurons directly implies that it is easier to correctly classify a node to be a dendrite, while it is more difficult to include all dendrite nodes by the same classifier. Hence, this explains why the precision is higher (or lower) than the recall for dendrites (or axons) of simple neurons (Fig. 5(a1) and (b1)). On the other hand, in Fig. 4(a2), axon nodes have a wider distribution than dendrite nodes in complex neurons, explaining why the precision is lower (or higher) than the recall for dendrites (or axons) of complex neurons (see Fig. 5(a2) and (b2)).

In addition to the path length to the soma, we have also included the direct distance from a node to a soma as a feature (Appendix B and Fig. S2(b)). Besides, the ratio of direct distance to the path length reflects a global morphological feature of a given node: if the distance to a soma is close to the path length to a soma, the neuron branches are more straight in the real space. The path is more curved if this ratio is much smaller than one. This implied that the node is close to the soma in space with a long and curved neuronal branch in between. In Figs. 4(b1) and (b2), we show the distribution of axon and dendrite nodes in the space of normalized length to the soma and normalized direct distance to the soma. The distribution clearly indicates that most nodes are well-separated in such 2D space. In fact, feature ranking by XGB also reveals these two features as the most important features for the identification of nodal polarity.

Apart from the two soma features mentioned above, in Fig. 4(c1) and (c2), we also present the distribution of nodal polarity in the space of normalized length to the soma and the cluster curvature near a given node. We suggest that the polarity classification can be effectively enhanced by including curvature as one of the local features because visual inspection reveals that typically more dendrites (compared to the axon nodes) can be found in the regime of larger curvatures. Such effects look more significant in simple neurons than in complex neurons. However, if we use curvature or other local features alone, the performance of polarity classification cannot be as good as using the path length to the soma.

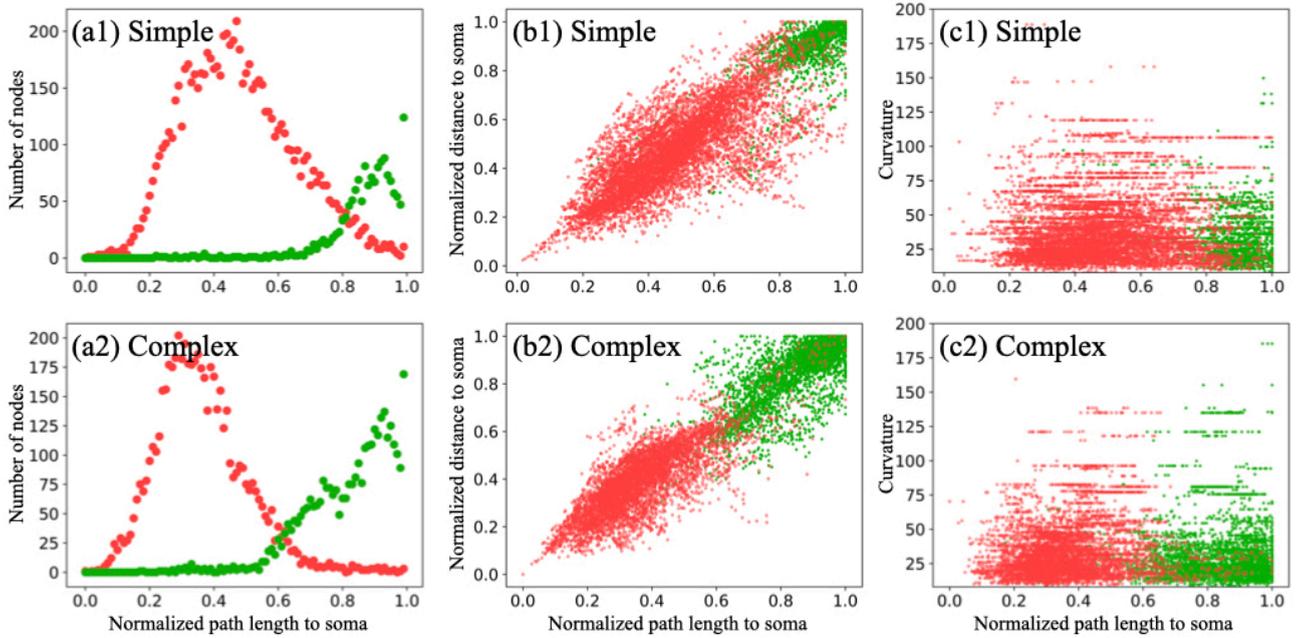

**Fig. 4 Feature distributions of axons and dendrites for all neurons in our dataset.** (a1) and (a2) show the distribution of axon and dendrite nodes along the normalized path length to soma, for simple and complex neurons, respectively. (b1) and (b2) display the nodal distribution in terms of the normalized path length and the normalized distance to the soma. (c1) and (c3) show the nodal distribution in terms of the normalized path length to the soma and the curvature of the associated cluster. Green and red dots represent axon and dendrite nodes, respectively. Details of curvature calculations are described in Appendix B.

The importance of each feature can also be obtained from the feature ranking calculation of XGB (however, this function is not available in DNN). This can also be obtained by comparing the overall accuracy after systematically removing certain features during the training. Our result suggests the top six features for the determination of nodal polarity: (1) unnormalized path length to the soma, (2) normalized path length to the soma, (3) unnormalized distance to the soma, (4) normalized distance to the soma, (5) curvature of the associated cluster, and (6) aspect ratio of the associated cluster. Other features are less important but can still contribute to the overall performance of NPIN. These results also confirm that local features are secondary factors for the determination of nodal polarity.

### Identification Results of Model I: Using Both Soma Features and Local Features

To present the results of polarity identification by NPIN, we start from Model I by using both soma features and local features for the whole dataset (with both simple and complex neurons). Figs. 5(a1)–(a3) show the confusion matrix of Model I based on XGB and the associated precision/recall table for the polarity of terminals in simple neurons, complex neurons, and all neurons, respectively. Figs. 5(b1)–(b3) show the results of Model I but based on DNN for comparison. Fig. 5(c) presents the definition of the confusion matrix and explains how the precision and recall are calculated for axons and dendrites. Because the final result is a binary classification of terminal polarity, the dividing nodes are not included either in the training data or in the test data.

From results shown in Figs. 5(a3) and 5(b3), we discover that NPIN is a very powerful classifier with an overall accuracy of 96%. This is achieved by including both Soma Features and Local Features. The model is trained and applied on both simple and complex neurons. According to our results, in general, the precision and recall for the polarity identification of dendritic terminals are better than those of axons by 3%–8%. One possible reason is that the total number of dendrite terminals is approximately three times more than the number of axon terminals, providing more training data that may increase the precision.

Comparing the confusing matrices for simple neurons (Figs. 5(a1) and 5(b1)) and complex neurons (Figs. 5(a2) and 5(b2)), we can observe similar performance of XGB and DNN on simple and complex neurons: The accuracy for simple neurons is higher than that of complex neurons by 1.2% for XGB (compare Figs. 5(a1) and 5(a2)), while it becomes 0.8% if calculated by DNN (compare Figs. 5(b1) and 5(b2)).

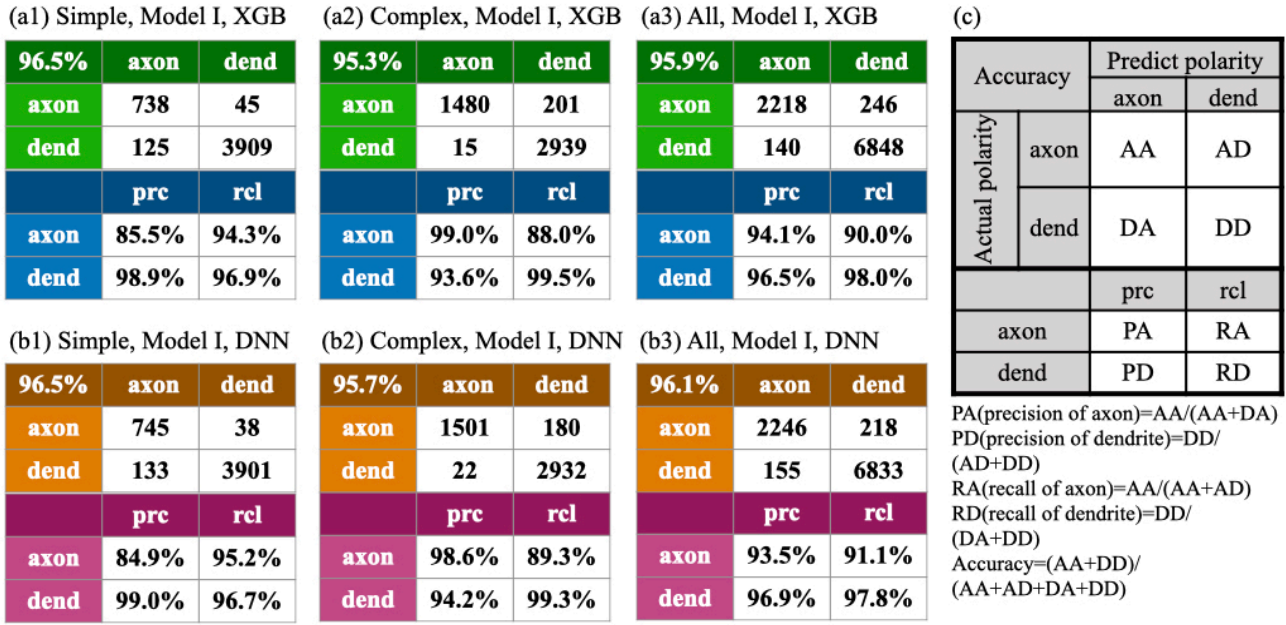

**Fig. 5 Performance of NPIN with Model I, where both Soma Features and Local Features are used.** (a1)–(a3) are the confusion matrix and precision/recall table of the terminal polarity, based on the XGB algorithm for simple, complex, and all neurons, respectively. (b1)–(b3) are the same as in (a1)–(a3) but calculated by the DNN algorithm. (c) defines the confusion matrices shown in this figure. In the upper part of the table, each row indicates the actual polarity, and each column indicates the polarity predicted by NPIN. The lower part of the table displays the precision and recall of axonal and dendritic terminals. Precision and recall are defined in the equations below (c).

However, such similar accuracy of polarity identification for simple and complex neurons is surprising, because complex neurons have more than two clusters. Therefore, the polarity of middle clusters cannot be easily identified according to its relative distance to soma. There are also various kinds of complex neurons (see Fig. 2, for example), which may also have an axon cluster close to the soma. As a result, a naïve comparison of the path length to the soma should not work well for a complex neuron. Hence, it is reasonable to believe that, in our NPIN, the contribution of soma features to simple and complex neurons should be different from the contribution of local features. To understand how this result is related to the feature selection in various types of neurons, in the following section, we demonstrate the performance of NPIN with different feature selections.

## Identification Results of Model II: Using Soma Features Only

To clarify the role of Soma Features and Local Features in the identification of neuronal polarity, we additionally use Model II, which is trained by using Soma Features only. The model is trained using the same protocol as in the previous section. The results are shown in Fig. 6.

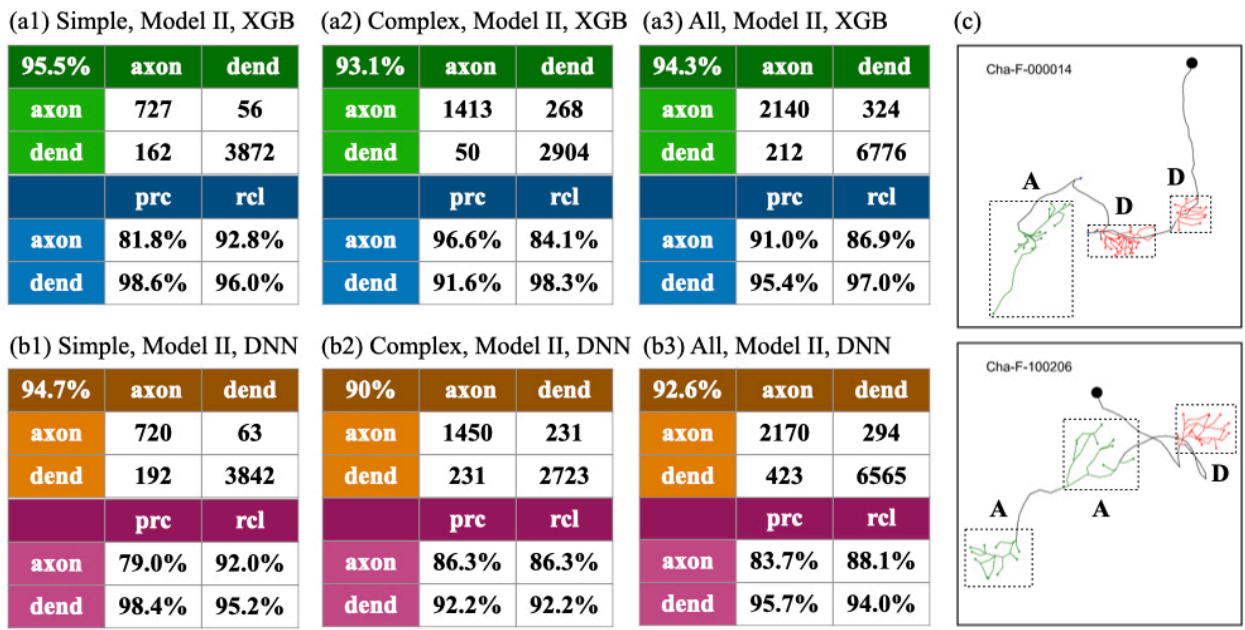

**Fig. 6 Performance of NPIN using Model II, where only Soma Features are included.** (a1)–(a3) show the results for simple neurons, complex neurons, and all neurons, respectively, using the XGB algorithm. (b1)-(b3) are the same as (a1)–(a3) but for the DNN algorithm. (c) shows two similar complex neurons, where middle clusters have opposite polarities. The cluster labeled by A/D is axons/dendrites.

According to Fig. 6, when using Soma Features only, we find that the overall accuracy drops to 95.5% (94.7%) for simple neurons, and 93.1% (90.0%) for complex neurons, respectively, if using XGB(DNN) algorithms. The performance on all neurons, as shown in Fig. 6(a3) and (b3), is between those of the simple and complex neurons, as expected.

Several important conclusions can be made. First, the overall accuracy of Model II is lower than for Model I (compare Fig. 6(a3) with Fig. 5(a3) for XGB and compare Fig. 6(b3) with Fig. 5(b3) for DNN). However, the difference is only 1.6% for XGB, while it is 3.5% for DNN. This means that the contribution of local features, which exists in Model I but not in Model II, is more significant for DNN than XGB. Second, if we compare the results for simple and complex neurons, we can see that the influence of local features is much more significant for complex neurons than for simple neurons. For example, for XGB, we find that the accuracy decreases by 1% only in simple neurons (compare Fig. 5(a1) and Fig. 6(a1)), while it decreases by 2.3% for complex neurons (compare Fig. 5(a2) and Fig. 6(a2)). These two values become 2.2% and 5.7%, respectively, for DNN. This clearly implies that missing Local Features in Model II are more important for complex neurons rather than simple neurons. The most obvious reason is that complex neurons have more than two clusters and, therefore, the simple application of soma features could not provide enough information for the identification of polarity. As an example, Fig. 6(c) shows two types of complex neurons, where the middle clusters have different polarities. These middle clusters are difficult to classify by Soma Features only. As a result, we conclude that local features are crucial for the polarity identification of the middle clusters in complex neurons and DNN algorithm may be more sensitive to these differences than XGB.

## Comparison of Models I, II, and III for Complex Neurons

In this experiment, to investigate how NPIN works with complex neurons and to examine its relationship with local features, we focus on complex neurons only: no simple neurons are included in either training data or test data. Three models are used for comparison: Model I (with both Soma Features and Local Features), Model II (with Soma Features only), and Model III (with Local Features only). Because the influence of Local Features is more significant in DNN than in XGB (see above), here we will apply DNN algorithm only for simplicity.

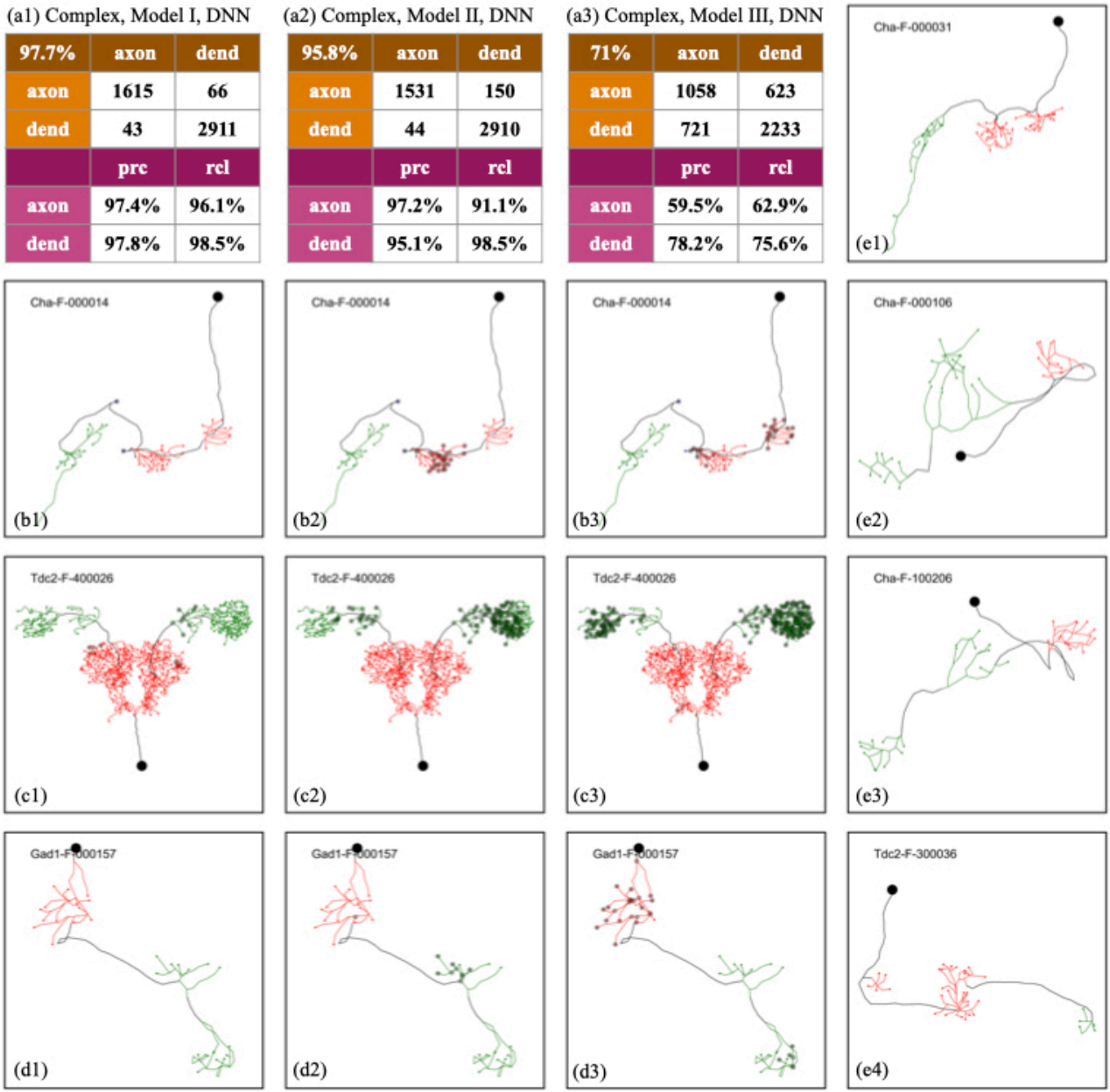

**Fig. 7 Performance of NPIN with DNN algorithm for complex neurons in three different models.** (a1)–(a3) are the confusion matrix and precision-recall table for the terminal polarity for Model I (with both Soma Features and Local Features), Model II (with Soma Features only), and Model III (with Local Features only), respectively. (b1)–(b3) display the same complex neuron with polarity classification using Model I, Model II, and Model III, respectively. Filled gray circles indicate the terminals of incorrect classification. (c1)–(c3) and (d1)–(d3) are the same as in (b1)–(b3) but with two different complex neurons. (e1)–(e4) are four different complex neurons, where polarities are classified by Model I with 100% accuracy by DNN algorithm.

According to the results shown in Fig. 7(a1)–(a3), the accuracy of classification is the best for Model I and slightly reduces for Model II, but it significantly drops for Model III (which uses Local Features only). This result indicates that, without any information on its relative distance to the soma, Local Features alone for a given node perform poorly in polarity identification but are not completely useless (with 71% accuracy, see Fig. 7(a3)). Indeed, we find that the inclusion of local features plays a complementary role in polarity identification, especially for the middle clusters of complex neurons. More precisely, by comparing Fig. 7(a2) to Fig. 7(a1), we find that local features can significantly reduce the number of incorrect identification for axons (upper right corner of the confusion table, from 150 to 66); hence, the number of correctly identified axons is increased.

In Fig. 7 (b1)–(b3), (c1)–(c3), and (d1)–(d3), we show three representative complex neurons with three or more clusters of terminals. Fig. 7 (b1), (b2), and (b3) show the same neuron with polarities identified by Model I, Model II, and Model III, respectively. Fig. 7 (c1)–(c3) and (d1)–(d3) show similar information but for another two neurons. The results obtained by using Local Features alone (Model III) are not satisfactory: some axon clusters with larger curvatures may be incorrectly classified as dendrites (see, for example, two axon clusters in Fig. 7(c3)). Moreover, some dendrite clusters with divergent branches may be incorrectly classified as axons (see, for example, the dendrite cluster in Fig. 7(d3)). Using Soma Features only (Model II), on the other hand, provides a much better result (with an accuracy of 95.8%), because clusters that are closest to or farthest from the soma are identified as dendrites or axons, respectively. However, as we see in Fig. 7(c2), (d2), and (e2), the middle clusters (defined from their distance to soma) of these complex neurons cannot be identified easily by Model II (with Soma Features only), because their relative distance to the soma is not well-defined compared to the other clusters.

As a summary, we find that the accuracy to classify the polarity of middle clusters in a complex neuron can be significantly enhanced after combining Soma Features and Local Features in Model I. More examples of complex neurons with correct polarity identification by Model I are shown in Fig. 7(e1)–(e4).

## Application to the Blowfly

In principle, our NPIN, trained on the *Drosophila* brain neurons, can also be applied to the polarity identification of other species, if the training data is replaced by the neurons of that species. However, the number of publicly available neuronal data samples of other species with identified polarity is much less than that of *Drosophila*. Therefore, such application may not be practical. However, it is still instructive to see how our NPIN, trained by *Drosophila* neurons, can be directly used for other species of insects, which should have similar morphological features as *Drosophila*. Here, we take the blowfly dataset from the Neuromorpho database (http://neuromorpho.org/) as an example. The database lists 19 blowfly neurons with labeled polarity. These data were generated by a different lab using a different reconstruction method from that of our *Drosophila* dataset.

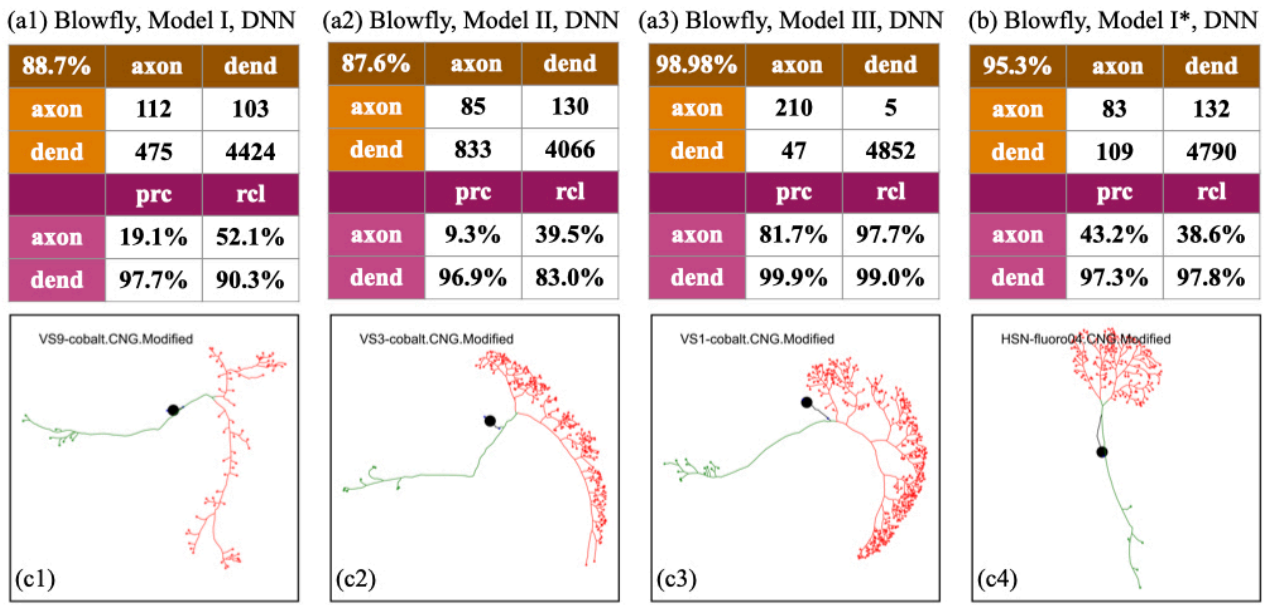

**Fig. 8 Performance of NPIN on blowfly brain neurons.** (a1)–(a3) are the confusion matrices and precision-recall tables for Model I, Model II, and Model III, respectively. The models are trained on 213 fruit-fly neurons in our dataset. (b) is the result for Model I but trained on blowfly neurons directly. (c1)–(c4) display four example skeleton structures of the blowfly neurons used in this test.

Fig. 8 shows the results of polarity identification for 19 blowfly neurons obtained by Model I, Model II, and Model III of NPIN, which is trained on 213 *Drosophila* neurons in our dataset. We find that Model I, using both soma features and local features, still provides a decent level of accuracy (83.4%). The main error stems from the pretty low precision and recall of the axons, which have much fewer terminal numbers than dendrites (dendrites: axons ratio = 22.8:1). Similar results are also observed for Model II, as shown in Fig. 8(a2).

However, a surprising result is obtained when using Model III, where only local features are included for the training on *Drosophila* neurons. The overall accuracy, as well as the precision and recall for both dendrites and axons, are very high (accuracy = 98.98%). This result is even better than that obtained by using the blowfly data for the training process (Fig. 8(b)). The results clearly indicate that, unlike *Drosophila*, where Local Features are only secondary factors compared with Soma Features, Local Features are the primary factors for the identification of neuronal polarity for blowfly neurons that we tested in the present study. This can also be observed from the skeleton structure of dendrite clusters in Fig. 8(c1)–(c4). Therefore, to apply NPIN (trained on *Drosophila* neurons) to neurons of other insects, it is necessary to provide not only Model I, but also Model II and Model III, to maximize the range of applications.

## Summary of Results

We summarize the results of the present study in Fig. 9 by showing the accuracy of NPIN in all test conditions including three models (Model I: all features, Model II: Soma Features only, Model III: Local Features only) and three types of test data (simple neurons, complex neurons, and all neurons). For simplicity, we only display the results using the DNN algorithm.

As explained above, the overall accuracy cannot reflect the complete information on model performance, especially when the numbers of dendrites and axons are highly imbalanced. To generate a reliable ML model, we suggest that the precision and recall for both axons and dendrites have to be larger than 50%, or, in other words, we have more correctly identified terminals than incorrect ones. We put stars "*" in Fig. 9 to mark those results that do not meet these criteria.

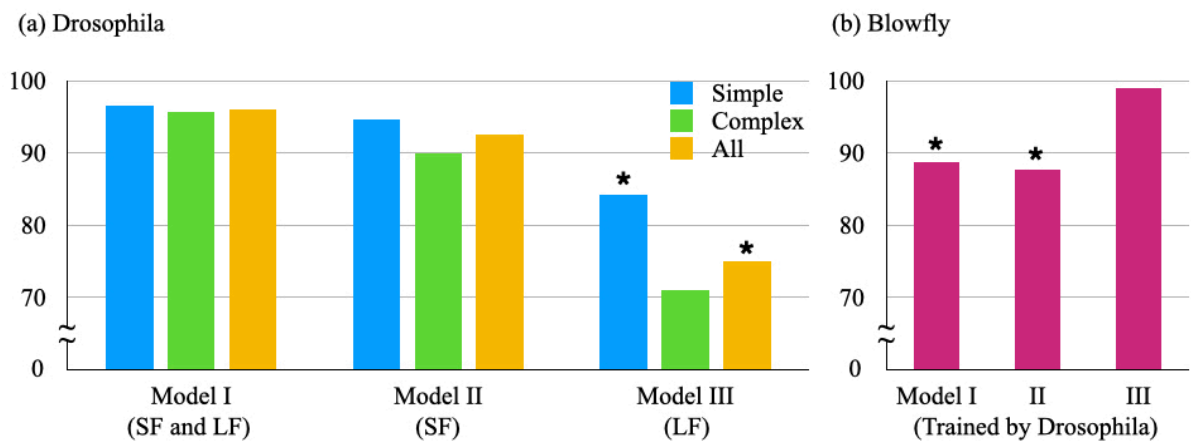

**Fig. 9 Summary of NPIN accuracies in all test conditions using the DNN algorithm.** (a) shows the results for Model I (with both Soma Features and Local Features), Model II (with Soma Features only), and Model III (with Local Features only), for three types of test data: simple neurons, complex neurons, and all neurons, respectively. (b) shows the results for the same models but with the blowfly neurons (trained by our *Drosophila* dataset). Results with precision or recall of less than 50% are indicated by "*" (see the text).

# Discussion

## Comparison of NPIN and SPIN

A previously developed machine-learning-based method, SPIN (described in the introduction), has identified the polarity of insect's neurons with an overall accuracy 84%–90% (Lee et al. 2014). SPIN starts by identifying clusters of neuronal arbors in each neuron and then classifies the polarity of each cluster according to its geometric structure and distance to the soma. As a result, terminals in a cluster are all classified as having the same polarity. However, this approach has two challenges. First, a cluster might not be easily identified for neurons with complex morphology, and incorrect clustering could lead to a large number of incorrectly classified terminals. Second, the number of available clusters may not be sufficient to achieve good training results because each neuron has only a few clusters. Due to these issues, SPIN often failed to classify part of or even all terminals of a neuron if its arbors were not clustered correctly.

The proposed NPIN avoids these issues by adopting node-based rather than cluster-based classification. To compare the performance of SPIN and NPIN, we examine the results of the polarity identification by SPIN on the same 213 neurons we used here (Huang et al. 2019). We find that, among these 213 neurons, only 79 neurons are fully identified (i.e., without any "non-classified" terminals), 120 neurons are partially predicted (i.e., some clusters cannot be identified), and 14 neurons cannot be predicted. Among 9452 terminals of these 213 neurons, there are 1207 unclassified terminals and 8247 classified terminals. Within the SPIN-classified terminals, 8038 terminals are correctly identified for their polarities. Therefore, the overall accuracy of SPIN is 85.04% only if we consider all terminals in the dataset, while it could be 97.49% if we considered only classified terminals.

We emphasize that, in the present study, we develop a completely different approach by identifying the polarity of each node, which can be unambiguously defined in the skeleton structure of each neuron, with a nodal polarity also well-defined through the polarity of terminals (see Fig. 3). Such node-based feature extraction, therefore, takes advantage of that the number of nodes is much larger than the number of clusters in each neuron. It can achieve a much higher accuracy (>96%) for the whole dataset (213 neurons and 9452 terminals) after including the spatial correlation. Therefore, NPIN outperforms SPIN in the polarity identification.

## Neurons with Low Accuracy

To examine the performance of our NPIN, we investigate those neurons not identified well in their polarity. As described in the Results section, we could obtain this information by randomly selecting 150 neurons (100 for training, 25 for validation, and 50 for testing) out of the 213 neurons in the dataset for each training/test process and then repeating it for 20 rounds. As a result, each neuron can be tested (by different models trained by other neurons) for 4–5 times on average, and their polarity identification results can be obtained by averaging their probabilities before relabeling. The final results calculated by the DNN model are shown in Appendix E. Within these 213 neurons, the terminal polarity of 166 neurons is identified with 100% accuracy. Only 14 simple neurons and 33 complex neurons are not fully identified. Concentrating on those neurons with a lower accuracy (say below 85%), we find only 5 simple neurons and 24 complex neurons left.

When looking into the skeleton structures of these neurons with a lower accuracy, we find the following features of these neurons: Simple neurons have a very similar distance for axon clusters and dendrite clusters to the soma, and the number of dendrite terminals is much larger than the number of axon terminals. The former makes it difficult to distinguish axons from dendrites, while the latter could confuse NPIN by mispredicting all terminals to be dendrites (as a result, the precision and recall of axons are both small). For complex neurons, the incorrectly identified terminals usually appear in the middle clusters, as one may expect. However, the most complex neurons have been correctly predicted by NPIN with a very high accuracy (91 of the 124 complex neurons are identified with 100% accuracy). In our node-based feature extraction, it is challenging to correctly identify the clusters of fewer terminals or nodes, because their local features are less representative of their local morphology. Therefore, this results indicate that we could not exclude the possibility of finding a better way to define local features (less dependent on the number of terminals in the same clusters) to enhance the results of polarity identification in future work.

## Other Types of Neurons

In the present study, we have tested 213 simple and complex neurons that cover a wide area of the *Drosophila* brain. However, this number is still far less than the total number of neurons (approximately 13.5K). Therefore, there could be other types of neurons with polarity-specific morphological features, which are very different from what we have addressed in this study. For example, the dendrites and axons of some local neurons are co-localized in the same cluster of arbors, and some projection neurons develop axonal clusters that are closer to the soma than the dendritic ones. We will include more morphologically distinct neurons into the training set, once their experimentally verified polarities are available. Therefore, although more training data are necessary when applying our NPIN for the polarity identification of the whole *Drosophila* brain, the present work at least demonstrates the possibilities to have a high precision identification through the node-based feature extraction in NPIN. We believe that future versions of NPIN, after including more types of neurons in the training data, will provide a much wider range of applications.

Finally, a large set of electronic-microscopy images (the EM dataset) of the *Drosophila* brain has recently been released (C. S. Xu et al. 2020). This dataset includes identified polarities, and it can be potentially used as training data for NPIN. However, after careful examination of the dataset, we discovered two major differences in the morphological characteristics between the two datasets: (1) the neuronal skeletons in the EM dataset exhibit much more details, e.g., a larger number of short terminal branches than what have been found in the fluorescent images in the present study. (2) Some neurons in the EM dataset have incomplete tracing or discontinuous branches. These issues prevent us from directly using the EM dataset for training. We suggest that heavy preprocessing is required before NPIN can utilize the EM dataset, and this will be addressed in future work.

## Conclusion

In this study, we have developed NPIN, a completely new ML model to identify the polarity of projection neurons in a *Drosophila* brain with high precision (>96%). This result was achieved due to three major contributions: node-based feature extraction, separation of Local Features from Soma Features, and implementation of spatial correlations between nodal polarities. In the experiments, we systematically compare the results of different models for various types of neurons. We demonstrate that, apart from Soma Features, Local Features are the secondary factors to determine the neuronal polarity. Local Features can significantly improve the polarity identification, especially for the middle clusters of complex neurons, which cannot be well-identified by using Soma Features only. Besides the *Drosophila* neurons, we show that NPIN can also be applied to identify the neuronal polarity of other insects, such as the blowfly. As a result, we believe that the development of NPIN and its applications is an important step toward the determination of signal flows in complex neural networks.

## Information sharing statement

The NPIN software package contains data of sample neurons with skeletal data available from the FlyCircuit database (http://www.flycircuit.tw/). We also provide an online version of NPIN to be used or tested by other research groups at the following address: https://npin-for-drosophila.herokuapp.com/

**Acknowledgments:** This work is supported by the Ministry of Science and Technology grant (MOST 107-2112-M-007-019-MY3) and by the Higher Education Sprout Project funded by the Ministry of Science and Technology and the Ministry of Education in Taiwan. We thank the National Center for Theoretical Sciences and Brain Research Center in the National Tsing Hua University for providing full support in the interdisciplinary collaboration. We thank the National Center for High-Performance Computing for providing the FlyCircuit database. We appreciate Prof. Che-Rung Lee for providing computational resources, Chiau-Jou Li and Yi-Ning Juan for helping with the website, Dr. Yu-Chi Huang and Prof. An-Shi Chiang for helpful discussion about the neuronal datasets.

**Conflict of interests:** The authors declare that they have no conflict of interests.

# Appendix

## A. Generation of a Reduced Tree from a Level Tree

    The level tree defined in this paper (see *Standardized Representation: Level Trees and Reduced Trees* and Fig. 3) has contained enough neuronal information for the determination of the nodal polarity. However, there are in general so many levels and nodes are involved that certain important features, like "clusters", cannot be defined easily from a computational point of view. In order to have a better definition of Local Features, which are more related to the geometric nature of the domain a node belongs to, in this paper we use a systematic method to obtain a "cluster" by trimming less important branches and then keeping the main trunk of a neuron. Those nodes left in such reduced trees on the trunks are then defined to be "the heads of clusters" (see for example, the node $i$ in Fig. S2(b) below). We then use their spatial information to define "clusters" and calculate Local Features.

    The way we trim less important branches is described as following steps: First, we calculate the number distribution of the whole "leaves", which are the branches to connect each terminal (see Fig. S1(a)) and their upper level nodes for a given neuron, as a function of their path lengths. From this distribution, we could determine the characteristic length, which has the largest number of "leaves" than other lengths. Second, we trim those "leaves" if their length is smaller than the characteristic length mentioned above. As a result, there will be new leaves coming out for new "terminals" after such trimming in the first round. Third, we repeat the calculation of length distribution again and trim those "leaves" of relatively shorter path lengths. Finally, such a trimming process is terminated when no more "leaves" of path lengths shorter than the characteristic length in the distribution, or when the total number of levels in the reduced tree reaches five (which is a convention we choose according to experience). In the inset of Figs. S1(a) and S1(b) we show the skeleton structure of a complex neuron before and after such reduction for comparison. The difference between these two figures give us the information of "clusters". The level tree and reduced tree representation are then obtained from these two skeleton structures respectively.

    It is easy to see that this is an efficient and systematic way to determine "clusters" from the level tree, while there are certainly also other methods to define clusters (see, for example, Lee et al. 2014). However, we have to emphasize that the identification of nodal polarity should not be sensitive to the details of these definitions, because we use these "clusters" to calculate morphological features, such as aspect ratio and curvature (see Appendix B), instead of using their detailed information directly. The identification of nodal polarity is determined not only by using Local Features and Soma Features, but also by implementing spatial correlation of polarities between neighboring nodes.

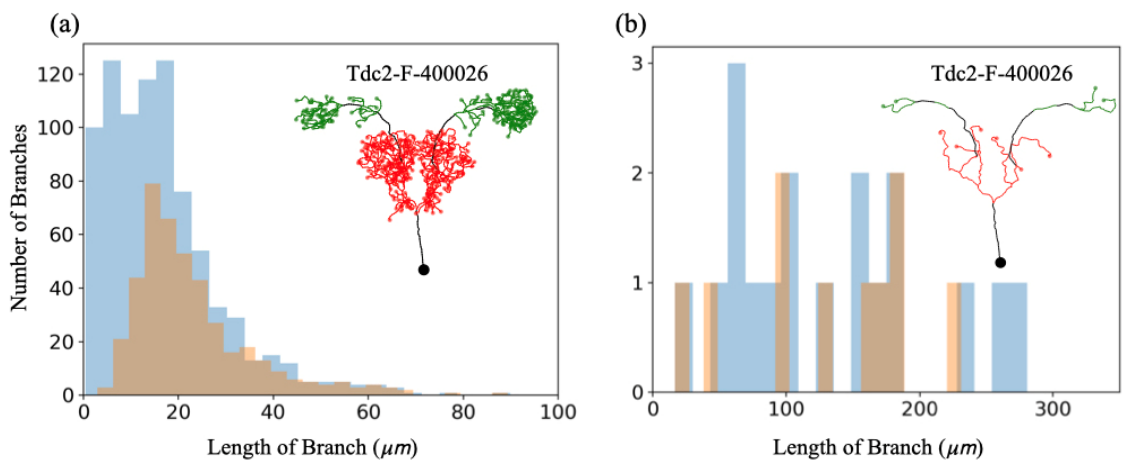

**Fig. S1** (a) shows the distribution of path length of all branches (blue) and all "leaves" (orange) of a given neuron. The threshold of trimming length is determined by the maximum value of the distribution for the "leaves". In this case, it is around 15 $\mu$m in the beginning. (b) shows the final distribution of all branches and "leaves" after repeating the trimming process. Since shorter and outer branches (leaves) are mostly trimmed, the final branches (after combining two or more branches) also become longer. The insets in

(a) and (b) show the skeleton structure before and after such a process. The level tree and reduced tree representations (shown in Fig. 3) are obtained from these two skeleton structures respectively.

## B. Calculation of Local Features: Curvature and Aspect Ratio

In this paper, we define two types of features for the polarity identification, Soma Features and Local Features respectively. It is easier to describe them in a skeleton diagram of neurons as shown in Figs. S2: For a given node, which is labeled as *i* here, Soma Features contain four values: (1) path length (along the neuron branch) to soma, $l_{si}$, see the thick line along *S-p-i*, (2) normalized path length to soma, $nl_{si} = l_{si}/L_s$, where $L_s$ is the largest path length to soma for a given neuron. (3) direct distance to soma, $d_{si}$, see the dashed straight line, *S-i*, and (4) normalized distance to soma, $nd_{si} = d_{si}/D_s$, where $D_s$ is the largest distance to soma for a given neuron. Note that we include both original path length/distance to soma and their normalized values because the former is to catch the possible size effects between different neurons, while the latter is for the comparison between nodes of the same neuron. These four features are defined to be Soma Features in this paper, because they are all related to the spatial information between nodes to soma.

Here we have to emphasize that Soma Features defined above contain not only the distance information for a node to soma, but also certain global shape information. The reason is that both path length to soma and direct distance to soma are included, and their ratio could imply how the neuronal truck and branches are curved in space between the given node and soma. In other words, our Soma Features provides more information than just the path length to soma. If using normalized path length to soma as the only feature for comparison, the accuracy of polarity prediction drops to 91.3% (by XGB), which is much lower than the accuracy (95.5%) obtained by using Soma Features defined above.

For Local Features, in this paper we include the following five values for a given node (see Fig. S2(b)): (1) path length to its parent node (in the upper level if in the level tree diagram), $l_{pi}$, see the thick line *p-i*, (2) normalized path length to its parent node, i.e. $nl_{pi} = l_{pi}/L_s$, (3) the curvature of the associated cluster, $c_i$, (4) the aspect ratio of the associated cluster, $ar_i$, and (5) the ratio of length, $rl_i = Min(l_{ij}, l_{ik})/(l_{ij} + l_{ik})$, which is to measure the relative differences between the lengths to its two children nodes. In Fig. S2(c), we show how to calculate the curvature and aspect ratio of a given "cluster", which is obtained from the definition of our reduced trees (see Appendix A). Note that all the Local Features are related only to the path length or shape structure of a given node with its neighboring nodes, and contains *no* information about its relationship to soma at all.

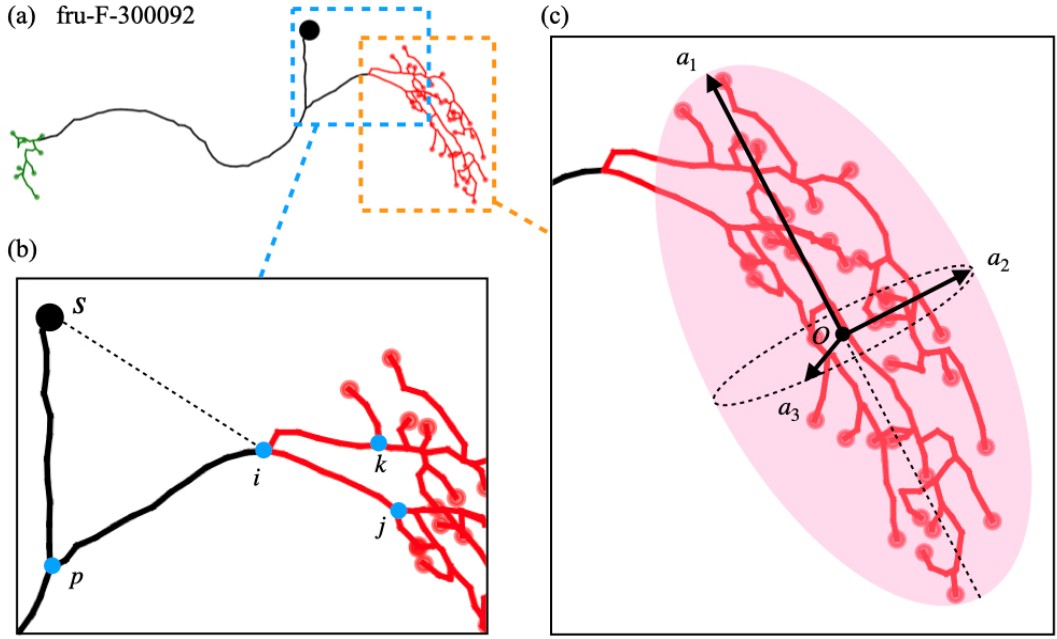

**Fig. S2** (a) shows a typical skeleton structure of a neuron in our dataset. In (b), we zoom in the neighboring skeleton structure of a given node, which is labeled by the index *i*. The path length to soma, direct distance to soma and the path length to other lower level nodes are denoted (see the text for details). In (c), we show the local morphological features of a given node inside the cluster. The curvature feature is defined by the ratio of total neuronal path length inside the cluster to the size of the cluster. This value indicates how much the neuronal branches are packed inside the cluster. The aspect ratio is to calculate the ratio of the longest length and the shortest length of a 3D elliptic surface, obtained according to the calculation of moment of inertia for this cluster. The details of calculation are described in the text.

In order to incorporate local morphological information for each node of a neuronal skeleton, two local features, curvature and aspect ratio, are defined for the "cluster", where a given node belongs to. The definition of "cluster" in this paper has been described in Appendix A and Fig. S1 above. To calculate the aspect ratio of such "cluster", we assume each node (including terminals) inside the cluster is a point of the same "mass", and therefore the moment of inertia with respect to their center of mass is given by the following two-rank tensor:

$$I = \begin{bmatrix} I_{xx} & I_{xy} & I_{xz} \\ I_{yx} & I_{yy} & I_{yz} \\ I_{zx} & I_{zy} & I_{zz} \end{bmatrix}$$

where each element can be calculated from

$$I_{xx} = \sum_{k \in cluster}(y_k^2 + z_k^2), I_{yy} = \sum_{k \in cluster}(x_k^2 + z_k^2), I_{zz} = \sum_{k \in cluster}(x_k^2 + y_k^2)$$

$$I_{xy} = I_{yx} = -\sum_{k \in cluster} x_k y_k, \; I_{xz} = I_{zx} = -\sum_{k \in cluster} x_k z_k, \; I_{yz} = I_{zy} = -\sum_{k \in cluster} y_k z_k$$

Here $x_k$, $y_k$, and $z_k$ are the position coordinates for each node (with a dummy index, $k$) relative to their center of mass point, $O$, which can be calculated easily (see Fig. S2(c)).

After diagonalization of this moment of inertia tensor, the obtained eigenvectors are orthogonal to each other and can be denoted to be $a_{1,2,3}$ as the three principal axes. Therefore, the aspect ratio of the cluster can be defined as $r \equiv Max[a_i/a_j]$ (see Fig. S2(c) as an example). After finding the length scale of the associated cluster, we can further define "curvature" of such cluster as, $c \equiv (total\ pathlength\ in\ the\ cluster)/(a_1 a_2 a_3)^{1/3}$, which indicates how tightly the nerve branches are packed inside such a cluster.

**C. Machine Learning Algorithms**

In this paper, we have used two machine learning algorithms, XGB and DNN. XGB uses gradient boosting methods based on a rule-based algorithm in order to optimize the accuracy in many-different tasks (Bekkerman 2015; Chen and Guestrin 2016). It belongs to so-called transparent models, because the algorithm could automatically evaluate the weighting of each feature and decide the best arrangement for each decision tree (See Fig. S3(a)). As a result, XGB could also provide feature ranking to evaluate the importance of these input features. In our NPIN, the XGB algorithm is designed with default hyper parameters from XGBoost Python Package[1]: the learning rate is 0.1, maximum depth of each tree is 3, and the number of trees is 100.

DNN was an algorithm originally designed to mimic the architecture of the human neural networks, but became one of the most general and powerful algorithms in the field of Artificial Intelligence (Deng and Yu 2014; LeCun et al. 2015; Schmidhuber 2015). Its common structure is composed of one input layer and one output layer, in between are multiple hidden layers (see Fig. S3(b)). In each layer, there are a number of interconnected nodes, or called "artificial neurons", which receive inputs from "artificial neurons" in the previous layer and supply outputs to others in the next layer. Each node performs a weighted sum computation on the values it receives from the input and then generates an output after a nonlinear transformation function on the summation. In our NPIN, the DNN algorithm has 9 features in the input layer (four Soma Features and five Local Features), propagated directly to the two fully connected hidden layers. Both of them contain ten artificial neurons and apply sigmoid function as activation functions. The output layer has two artificial neurons representing the probability of being axon or dendrite, using Focal Loss function (T.-Y. Lin et al. 2018) and Adam optimizer (Kingma and Ba 2017) for the training processes.

Therefore, we can see that XGB and DNN are complementary algorithms in many different perspectives: First of all, XGB is a transparent model, which could provide certain results for feature ranking, while DNN is known not explainable due to its highly non-linear function coupling. Besides, it is known that XGB does not need much data (as well as computational time) for the training process, while DNN usually needs a large amount of data and computational power. On the other hand, when the amount of training data is sufficiently large, DNN in general provides higher precision than transparent models, especially when the features are highly correlated to each other.

---

[1] https://hcho3-xgboost.readthedocs.io/en/latest/python/python_api.html

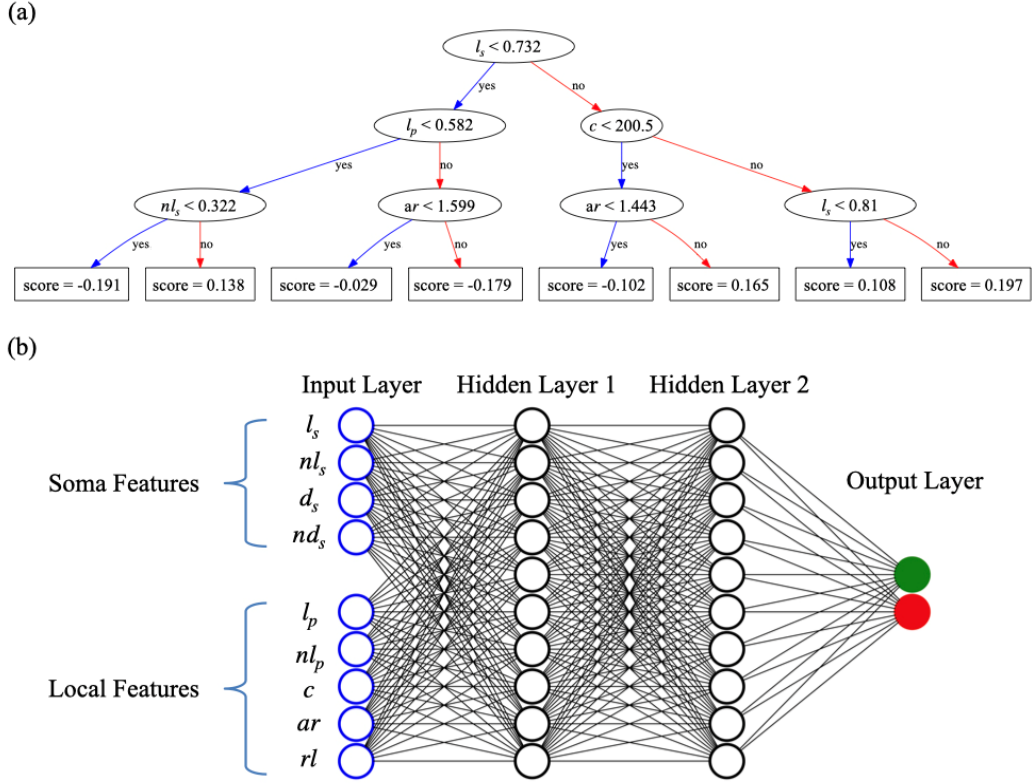

**Fig. S3** (a) shows a typical decision tree diagram, obtained by XGB with feature ranking. Each oval represents a criterion and the scores from all the decision trees are used for calculating the probability to be axon (higher score) or dendrite (lower scores) (Chen and Guestrin 2016). (b) shows the layer structure of our DNN model, where we use fully connecting networks between artificial neurons in the neighboring layers.

### D. Including Spatial Nodal Correlation through Polarity Relabeling

In the NPIN model, the accuracy of polarity classification can be further improved by imposing the spatial correlation between nodal polarities. This is because, except for some local neurons which are not included in our analysis, axonal or dendritic terminals in a neuron appear in clusters. However, in our XGB or DNN algorithms, the polarity of each node is independently classified without knowing polarities of neighboring nodes. The spatial correlation is then imposed by relabeling the polarities of nodes, which have low probability scores from the classification process, and the polarity of a relabeled node is determined by its neighboring nodes that have higher probability scores.

The basic relabeling procedures are following: First we relabel the polarity of all the nodes according to the nodal polarity in the lower level, see Fig. S4. In other words, although we could predict all the nodal polarity as well as the terminal polarity by XGB or DNN, the terminals of high probabilities to be axons or dendrites still have higher priority to determine the polarity of nodes in the upper levels. On the other hand, those terminals predicted to have lower probabilities to be either axons or dendrites will then be determined from nodes in the upper levels and its neighbors in the second stage. Finally, we define dividing nodes in the same way as we described for the data preparation (see Fig. 3(c)). This stage will not rewrite any polarity except to identify dividing nodes in order to make the level tree representation consistent with the original definition.

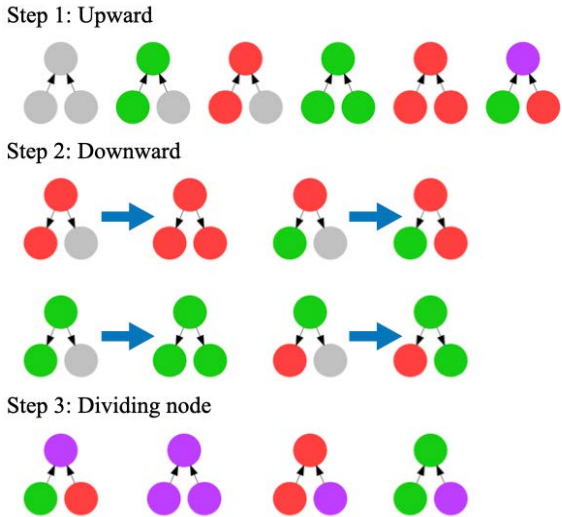

**Fig. S4** shows the three steps to relabel nodal polarity by including the spatial correlation between neighboring nodes. Grey nodes are nodes with low probability scores from the polarity classification, green nodes are classified axons, red nodes are classified dendrites and purple nodes are the dividing nodes. In Step 1, nodal polarity is determined by the polarity of nodes in the lower level, taking into account the existence of nodes with probabilities lower than the threshold (filled grey circles). In Step 2, grey circles are relabeled according to the results of its parent nodes in the upper level. In Step 3, we complete the relabeling process by identifying dividing nodes. The final results are consistent with the original definition of nodal polarity shown in Fig. 3(c), but we use the polarities of terminal nodes only to get the ground truth.

In order to determine the criteria for a node to enter the relabeling process, in Fig. S5(a) we plot the distribution of nodes based on the probability being an axon from the XGB classifier. Since it is a binary classification, a higher probability to be an axon must indicate a lower probability to be a dendrite, and vice versa. We can see that most nodes are classified to be either axon or dendrite with high probabilities (as shown by peaks on the two sides), but there are still a small portion of nodes with low probabilities. In Fig. S5(b1). We show the level tree of a given neuron with preliminary results given by XGB. One could see that although most nodal polarities are correctly identified, some are incorrectly identified. In Fig. S5(b2) we show the same level tree by removing the polarity labels of nodes which have probabilities being axons or dendrites below 0.75. In Fig. S5(b3), we show the final results of the level tree after relabeling (see Fig. S4): all nodes now are correctly classified after imposing spatial correlation through the relabeling process.

In Figs. S5(c) and S5(d) we show the comparison of the precision/recall table before and after such relabeling. It shows that the spatial correlation may significantly enhance the recall of axons by XGB, while it does not enhance that much for DNN. Since the spatial correlation is also strongly related to the distribution of neighboring nodes, the above result reflects the fact that, in a rule-based algorithm like XGB, morphologic features (curvature and aspect ratio) are relatively less addressed. However, spatial correlation can be learned to some degree through these morphologic features in DNN, making such post-relabeling less effective.

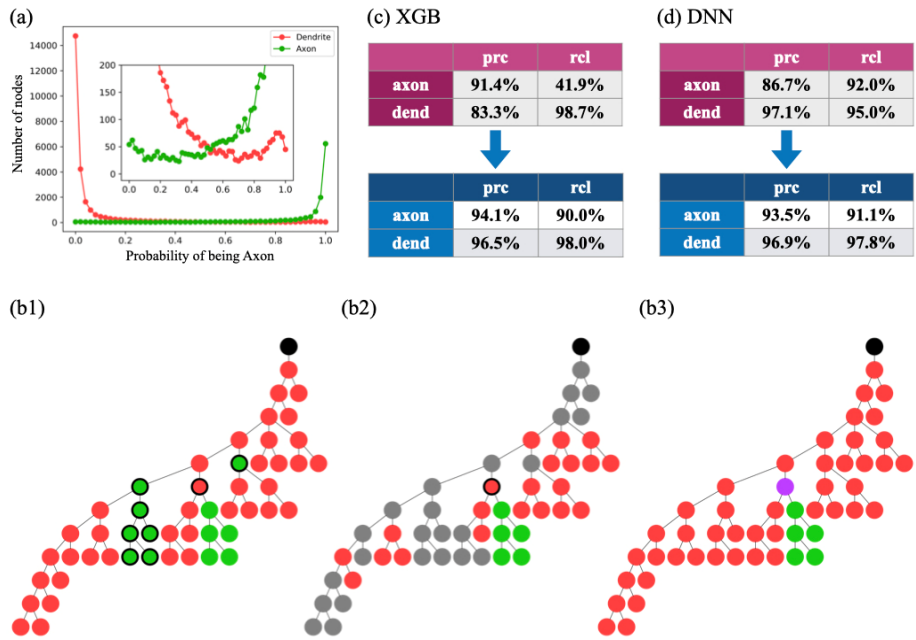

**Fig. S5** (a) shows the distributions of the probability being an axon as classified by XGB. The distributions are plotted separately for nodes that are actually axons (green) or dendrites (red) for all 213 neurons in our dataset. Note that this is a binary classification, so the sum of the probability to be dendrite and the probability to be axon has to be equal to one. (b1) is the level trees of the preliminary result given by XGB, where some nodes are incorrectly identified, as indicated by thick black circles. (b2) is the same level tree with removed polarity labels for some nodes (grey), if their probabilities be an axon and a dendrite are both below a threshold (=0.75). (b3) shows the final results after relabeling. All the axons and dendrites are correctly classified after imposing the spatial correlation of nodal polarity. (c) and (d) are the precision/recall tables provided by XGB and DNN, respectively for Model I (using both Soma Features and Local Features), applied on the whole dataset. The upper tables are the results before the relabeling process and the lower ones are after.

## E. Neuron Dataset for NPIN and Their Polarity Identification Results

Here we present the results of polarity identification by NPIN for all neurons in our dataset. It is obtained by randomly selecting 150 neurons (100 for training, 25 for validation, and 50 for testing) out of the 213 neurons in the dataset for each training/test process, and we repeat the same calculation for 20 rounds. As a result, each neuron can be tested (by different models trained by other neurons) for 4 or 5 times on average and the terminal polarities are identified by averaging their probabilities before final relabeling.

**Tabel S1. Polarity Identification of Simple Neurons in the Dataset**

| Neuron ID | Brain Region | | No. of Terminals | | Precision | | Recall | | Accuracy |
|---|---|---|---|---|---|---|---|---|---|
| | Axon | Dendrite | Axon | Dendrite | Axon | Dendrite | Axon | Dendrite | |
| 5HT1A-F-100004 | LH | AL | 18 | 64 | 1.00 | 1.00 | 1.00 | 1.00 | 1.00 |
| 5HT1A-F-100032 | LH | AL | 23 | 132 | 1.00 | 1.00 | 1.00 | 1.00 | 1.00 |
| 5HT1A-F-600006 | LH | AL | 11 | 29 | 1.00 | 1.00 | 1.00 | 1.00 | 1.00 |
| Cha-F-100132 | AOTU | MED | 2 | 55 | 1.00 | 1.00 | 1.00 | 1.00 | 1.00 |
| Cha-F-100165 | AOTU | MED | 3 | 28 | 1.00 | 1.00 | 1.00 | 1.00 | 1.00 |
| Cha-F-100352 | AOTU | MED | 2 | 20 | 1.00 | 1.00 | 1.00 | 1.00 | 1.00 |
| Cha-F-100414 | AOTU | MED | 2 | 38 | 1.00 | 1.00 | 1.00 | 1.00 | 1.00 |
| Cha-F-300281 | AOTU | MED | 6 | 104 | 1.00 | 1.00 | 1.00 | 1.00 | 1.00 |
| Cha-F-500233 | AOTU | MED | 4 | 59 | 1.00 | 1.00 | 1.00 | 1.00 | 1.00 |
| Cha-F-600158 | AOTU | MED | 2 | 36 | 1.00 | 1.00 | 1.00 | 1.00 | 1.00 |
| Cha-F-600239 | AOTU | MED | 3 | 48 | 1.00 | 1.00 | 1.00 | 1.00 | 1.00 |
| Cha-F-600263 | AOTU | MED | 4 | 63 | 1.00 | 1.00 | 1.00 | 1.00 | 1.00 |
| Cha-F-700223 | AOTU | MED | 1 | 43 | 1.00 | 1.00 | 1.00 | 1.00 | 1.00 |
| Cha-F-800020 | AOTU | MED | 2 | 69 | 1.00 | 1.00 | 1.00 | 1.00 | 1.00 |
| Cha-F-800097 | AOTU | LOB | 9 | 37 | 1.00 | 1.00 | 1.00 | 1.00 | 1.00 |
| fru-F-000012 | VMP | MED | 5 | 25 | 1.00 | 1.00 | 1.00 | 1.00 | 1.00 |
| fru-F-000230 | VMP | MED | 4 | 49 | 1.00 | 1.00 | 1.00 | 1.00 | 1.00 |
| fru-F-100050 | VMP | MED | 10 | 69 | 1.00 | 1.00 | 1.00 | 1.00 | 1.00 |
| fru-F-100096 | VMP | MED | 6 | 39 | 1.00 | 1.00 | 1.00 | 1.00 | 1.00 |
| fru-F-200085 | VMP | MED | 9 | 54 | 1.00 | 1.00 | 1.00 | 1.00 | 1.00 |
| fru-F-200101 | VMP | MED | 8 | 71 | 1.00 | 1.00 | 1.00 | 1.00 | 1.00 |
| fru-F-200153 | VMP | MED | 10 | 105 | 1.00 | 1.00 | 1.00 | 1.00 | 1.00 |
| fru-F-300092 | VMP | MED | 8 | 36 | 1.00 | 1.00 | 1.00 | 1.00 | 1.00 |
| fru-F-300111 | VMP | MED | 9 | 78 | 1.00 | 1.00 | 1.00 | 1.00 | 1.00 |
| fru-F-300120 | VMP | MED | 9 | 52 | 1.00 | 1.00 | 1.00 | 1.00 | 1.00 |
| fru-F-300130 | VMP | MED | 4 | 79 | 1.00 | 1.00 | 1.00 | 1.00 | 1.00 |
| fru-F-400057 | VMP | MED | 5 | 21 | .833 | 1.00 | 1.00 | .952 | .962 |
| fru-F-400058 | AOTU | LOB | 7 | 14 | 1.00 | 1.00 | 1.00 | 1.00 | 1.00 |
| fru-F-400209 | AOTU | LOB | 10 | 22 | 1.00 | 1.00 | 1.00 | 1.00 | 1.00 |
| fru-F-400341 | VMP | MED | 6 | 44 | 1.00 | 1.00 | 1.00 | 1.00 | 1.00 |
| fru-F-400366 | VMP | MED | 5 | 93 | 1.00 | 1.00 | 1.00 | 1.00 | 1.00 |
| fru-F-500119 | AOTU | LOB | 6 | 8 | 1.00 | 1.00 | 1.00 | 1.00 | 1.00 |
| fru-F-500257 | PB | CVLP,DMP | 12 | 83 | .857 | .928 | .500 | .987 | .922 |
| fru-F-500486 | AOTU | LOB | 2 | 12 | 1.00 | 1.00 | 1.00 | 1.00 | 1.00 |
| fru-F-500578 | VMP | MED | 7 | 55 | 1.00 | 1.00 | 1.00 | 1.00 | 1.00 |
| fru-F-600005 | VMP | MED | 8 | 31 | 1.00 | 1.00 | 1.00 | 1.00 | 1.00 |
| fru-F-700013 | VMP | MED | 8 | 28 | 1.00 | 1.00 | 1.00 | 1.00 | 1.00 |
| fru-F-700059 | VMP | MED | 10 | 39 | .909 | 1.00 | 1.00 | .974 | .980 |
| fru-F-700157 | VMP | MED | 4 | 74 | 1.00 | 1.00 | 1.00 | 1.00 | 1.00 |
| fru-F-800052 | VMP | MED | 10 | 91 | 1.00 | 1.00 | 1.00 | 1.00 | 1.00 |
| fru-F-800083 | VMP | MED | 6 | 64 | 1.00 | 1.00 | 1.00 | 1.00 | 1.00 |
| fru-F-900027 | VMP | MED | 8 | 67 | .364 | 1.00 | 1.00 | .788 | .811 |
| fru-F-900039 | VMP | MED | 10 | 71 | 1.00 | 1.00 | 1.00 | 1.00 | 1.00 |
| Gad1-F-000709 | AOTU | LOB | 21 | 31 | 1.00 | 1.00 | 1.00 | 1.00 | 1.00 |
| Gad1-F-000777 | AOTU | LOB | 15 | 44 | 1.00 | .955 | .846 | 1.00 | .964 |
| Gad1-F-200218 | AOTU | MED | 5 | 42 | 1.00 | 1.00 | 1.00 | 1.00 | 1.00 |
| Gad1-F-200389 | AOTU | LOB | 8 | 16 | 1.00 | 1.00 | 1.00 | 1.00 | 1.00 |
| Gad1-F-200780 | AOTU | LOB | 10 | 33 | 1.00 | 1.00 | 1.00 | 1.00 | 1.00 |
| Gad1-F-300536 | LH | AL | 10 | 44 | 1.00 | 1.00 | 1.00 | 1.00 | 1.00 |
| Gad1-F-400023 | AOTU | MED | 4 | 55 | 1.00 | 1.00 | 1.00 | 1.00 | 1.00 |
| Gad1-F-500071 | AOTU | LOB | 4 | 11 | 0.00 | 0.00 | 0.00 | 0.00 | 0.00 |
| Gad1-F-500088 | LH | AL | 13 | 42 | 1.00 | 1.00 | 1.00 | 1.00 | 1.00 |
| Gad1-F-500568 | LH | AL | 18 | 44 | 1.00 | 1.00 | 1.00 | 1.00 | 1.00 |
| Gad1-F-600331 | AOTU | LOB | 6 | 19 | 1.00 | .950 | .833 | 1.00 | .960 |
| Gad1-F-600560 | CAL | AL | 1 | 54 | 1.00 | 1.00 | 1.00 | 1.00 | 1.00 |
| Gad1-F-600676 | LH | AL | 11 | 157 | 1.00 | 1.00 | 1.00 | 1.00 | 1.00 |
| Gad1-F-700055 | AOTU | MED | 3 | 59 | 1.00 | 1.00 | 1.00 | 1.00 | 1.00 |
| Gad1-F-800354 | LH | AL | 5 | 27 | 1.00 | 1.00 | 1.00 | 1.00 | 1.00 |
| Gad1-F-800392 | AOTU | LOB | 5 | 32 | 1.00 | .889 | .200 | 1.00 | .892 |
| Gad1-F-900119 | AOTU | LOB | 9 | 20 | 1.00 | 1.00 | 1.00 | 1.00 | 1.00 |
| Gad1-F-900515 | LH | AL | 13 | 70 | 1.00 | 1.00 | 1.00 | 1.00 | 1.00 |
| TH-F-000048 | PB | CVLP,IDFP,VMP | 67 | 75 | .805 | 1.00 | 1.00 | .787 | .887 |
| Trh-F-300080 | LH | AL | 13 | 20 | 1.00 | 1.00 | 1.00 | 1.00 | 1.00 |
| Trh-F-400067 | LH | AL | 9 | 13 | 1.00 | 1.00 | 1.00 | 1.00 | 1.00 |
| Trh-F-500027 | LH | AL | 7 | 12 | 1.00 | 1.00 | 1.00 | 1.00 | 1.00 |
| Trh-F-500049 | LH | AL | 12 | 16 | 1.00 | 1.00 | 1.00 | 1.00 | 1.00 |
| Trh-F-500059 | LH | AL | 12 | 50 | 1.00 | 1.00 | 1.00 | 1.00 | 1.00 |
| Trh-F-500077 | LH | AL | 4 | 20 | 1.00 | 1.00 | 1.00 | 1.00 | 1.00 |
| Trh-F-600092 | LH | AL | 15 | 23 | 1.00 | 1.00 | 1.00 | 1.00 | 1.00 |
| Trh-F-600104 | LH | AL | 12 | 15 | 1.00 | 1.00 | 1.00 | 1.00 | 1.00 |
| Trh-F-700032 | LH | AL | 4 | 16 | 1.00 | 1.00 | 1.00 | 1.00 | 1.00 |
| VGlut-F-000259 | LH | AL | 38 | 138 | 1.00 | 1.00 | 1.00 | 1.00 | 1.00 |
| VGlut-F-000370 | LH | AL | 12 | 146 | 1.00 | 1.00 | 1.00 | 1.00 | 1.00 |
| VGlut-F-300584 | PB | CVLP,DMP | 19 | 83 | .667 | .819 | .105 | .987 | .814 |
| VGlut-F-400245 | CAL | AL | 7 | 47 | 1.00 | 1.00 | 1.00 | 1.00 | 1.00 |
| VGlut-F-400634 | LH | AL | 15 | 26 | 1.00 | 1.00 | 1.00 | 1.00 | 1.00 |
| VGlut-F-500092 | LH | AL | 11 | 27 | 1.00 | 1.00 | 1.00 | 1.00 | 1.00 |
| VGlut-F-500853 | VLP | LOB | 4 | 21 | 1.00 | 1.00 | 1.00 | 1.00 | 1.00 |
| VGlut-F-600669 | LH | AL | 11 | 51 | 1.00 | 1.00 | 1.00 | 1.00 | 1.00 |
| VGlut-F-600757 | LH | AL | 9 | 34 | 1.00 | 1.00 | 1.00 | 1.00 | 1.00 |
| VGlut-F-700021 | CAL | AL | 2 | 35 | 1.00 | 1.00 | 1.00 | 1.00 | 1.00 |
| VGlut-F-700072 | CAL | AL | 2 | 32 | 1.00 | 1.00 | 1.00 | 1.00 | 1.00 |
| VGlut-F-700163 | VLP | LOB | 2 | 39 | 0.00 | .917 | 0.00 | .564 | .537 |
| VGlut-F-700230 | VLP | LOB | 8 | 27 | .500 | .806 | .250 | .926 | .771 |
| VGlut-F-700402 | VLP | LOB | 10 | 12 | 1.00 | 1.00 | 1.00 | 1.00 | 1.00 |
| VGlut-F-800076 | VLP | LOB | 6 | 25 | .667 | 1.00 | 1.00 | .880 | .903 |
| VGlut-F-800224 | VLP | LOB | 11 | 67 | .688 | 1.00 | 1.00 | .917 | .930 |
| VGlut-F-800284 | LH | AL | 18 | 21 | 1.00 | 1.00 | 1.00 | 1.00 | 1.00 |
| VGlut-F-800305 | VLP | LOB | 5 | 17 | 1.00 | 1.00 | 1.00 | 1.00 | 1.00 |

**Tabel S2. Polarity Identification of Complex Neurons in the Dataset**

| Neuron ID | Brain Region | | No. of Terminals | | Precision | | Recall | | Accuracy |
|---|---|---|---|---|---|---|---|---|---|
| | Axon | Dendrite | Axon | Dendrite | Axon | Dendrite | Axon | Dendrite | |
| Cha-F-000014 | IDFP | FB,PB | 14 | 36 | 1.00 | 1.00 | 1.00 | 1.00 | 1.00 |
| Cha-F-000023 | FB,NO | PB | 14 | 6 | 1.00 | .556 | .714 | 1.00 | .789 |
| Cha-F-000031 | IDFP | FB,PB | 16 | 37 | 1.00 | .720 | .125 | 1.00 | .731 |
| Cha-F-000050 | PB | CCP,VMP | 9 | 39 | 0.00 | .804 | 0.00 | .974 | .787 |
| Cha-F-000098 | IDFP | FB,PB | 9 | 24 | 1.00 | 1.00 | 1.00 | 1.00 | 1.00 |
| Cha-F-000106 | FB,NO | PB | 23 | 11 | 1.00 | .478 | .478 | 1.00 | .647 |
| Cha-F-000112 | FB,NO | PB | 9 | 5 | 1.00 | 1.00 | 1.00 | 1.00 | 1.00 |
| Cha-F-000423 | EB,NO | PB | 34 | 6 | 1.00 | 1.00 | 1.00 | 1.00 | 1.00 |
| Cha-F-100032 | IDFP | FB,PB | 12 | 31 | 1.00 | 1.00 | 1.00 | 1.00 | 1.00 |
| Cha-F-100041 | IDFP | FB,PB | 19 | 67 | 1.00 | 1.00 | 1.00 | 1.00 | 1.00 |
| Cha-F-100065 | IDFP | FB,PB | 14 | 42 | 1.00 | 1.00 | 1.00 | 1.00 | 1.00 |
| Cha-F-100117 | PB | CCP,VMP | 9 | 49 | 0.00 | .833 | 0.00 | .957 | .804 |
| Cha-F-100206 | FB,NO | PB | 18 | 14 | 1.00 | .750 | .765 | 1.00 | .862 |
| Cha-F-200009 | IDFP | FB,PB | 14 | 20 | 1.00 | 1.00 | 1.00 | 1.00 | 1.00 |
| Cha-F-200013 | IDFP | FB,PB | 15 | 43 | 1.00 | .759 | .071 | 1.00 | .764 |
| Cha-F-200046 | IDFP | FB,PB | 8 | 18 | 1.00 | 1.00 | 1.00 | 1.00 | 1.00 |
| Cha-F-200068 | IDFP | FB,PB | 15 | 48 | 1.00 | .857 | .467 | 1.00 | .873 |
| Cha-F-200084 | IDFP | FB,PB | 10 | 18 | 1.00 | .810 | .556 | 1.00 | .846 |
| Cha-F-300072 | FB,NO | PB | 14 | 13 | 1.00 | 1.00 | 1.00 | 1.00 | 1.00 |
| Cha-F-300152 | FB,NO | PB | 13 | 7 | 1.00 | .389 | .154 | 1.00 | .450 |
| Cha-F-300160 | IDFP | FB,PB | 14 | 20 | 1.00 | .519 | .071 | 1.00 | .536 |
| Cha-F-400006 | FB,NO | PB | 19 | 8 | 1.00 | .636 | .789 | 1.00 | .846 |
| Cha-F-400012 | IDFP | FB,PB | 12 | 60 | 1.00 | .866 | .250 | 1.00 | .871 |
| Cha-F-400017 | IDFP | FB,PB | 11 | 50 | 1.00 | 1.00 | 1.00 | 1.00 | 1.00 |
| Cha-F-400025 | FB,NO | PB | 14 | 10 | 1.00 | 1.00 | 1.00 | 1.00 | 1.00 |
| Cha-F-400260 | FB,NO | PB | 17 | 11 | 1.00 | 1.00 | 1.00 | 1.00 | 1.00 |
| Cha-F-500009 | EB,NO | PB | 20 | 10 | 1.00 | 1.00 | 1.00 | 1.00 | 1.00 |
| Cha-F-500028 | IDFP | FB,PB | 21 | 41 | 1.00 | .784 | .421 | 1.00 | .814 |
| Cha-F-500046 | IDFP,PB | EB | 13 | 8 | .429 | 0.00 | .500 | 0.00 | .300 |
| Cha-F-500056 | CAL,LH | AL | 28 | 24 | 1.00 | 1.00 | 1.00 | 1.00 | 1.00 |
| Cha-F-500109 | IDFP | FB,PB | 9 | 52 | 1.00 | 1.00 | 1.00 | 1.00 | 1.00 |
| Cha-F-500285 | FB,NO | PB | 15 | 7 | 1.00 | 1.00 | 1.00 | 1.00 | 1.00 |
| Cha-F-600001 | IDFP | FB,PB | 14 | 47 | 1.00 | .836 | .357 | 1.00 | .850 |
| Cha-F-700086 | IDFP | FB,PB | 10 | 32 | 1.00 | 1.00 | 1.00 | 1.00 | 1.00 |
| fru-F-100063 | EB,NO | PB | 12 | 8 | 1.00 | 1.00 | 1.00 | 1.00 | 1.00 |
| fru-F-400276 | CAL,LH | AL | 17 | 54 | 1.00 | .964 | .882 | 1.00 | .972 |
| fru-F-500176 | CAL,LH | AL | 13 | 38 | 1.00 | 1.00 | 1.00 | 1.00 | 1.00 |
| fru-F-700239 | CAL,LH | AL | 5 | 11 | 1.00 | 1.00 | 1.00 | 1.00 | 1.00 |
| fru-M-400292 | CAL,LH | AL | 16 | 40 | 1.00 | 1.00 | 1.00 | 1.00 | 1.00 |
| fru-M-400387 | CAL,LH | AL | 12 | 11 | 1.00 | 1.00 | 1.00 | 1.00 | 1.00 |
| Gad1-F-000056 | FB,NO | PB | 13 | 4 | 1.00 | .500 | .667 | 1.00 | .750 |
| Gad1-F-000066 | FB,NO | PB | 18 | 8 | 1.00 | .381 | .278 | 1.00 | .500 |
| Gad1-F-000157 | FB,NO | PB | 17 | 14 | 1.00 | .800 | .812 | 1.00 | .893 |
| Gad1-F-000167 | CAL,LH | AL | 14 | 20 | 1.00 | 1.00 | 1.00 | 1.00 | 1.00 |
| Gad1-F-000172 | FB,NO | PB | 15 | 20 | 1.00 | 1.00 | 1.00 | 1.00 | 1.00 |
| Gad1-F-000671 | CAL,LH | AL | 11 | 21 | 1.00 | 1.00 | 1.00 | 1.00 | 1.00 |
| Gad1-F-100004 | IDFP | FB,PB | 8 | 23 | 1.00 | 1.00 | 1.00 | 1.00 | 1.00 |
| Gad1-F-100134 | CAL,LH | AL | 9 | 16 | 1.00 | 1.00 | 1.00 | 1.00 | 1.00 |
| Gad1-F-200375 | EB,NO | PB | 13 | 10 | 1.00 | 1.00 | 1.00 | 1.00 | 1.00 |
| Gad1-F-300027 | IDFP | FB,PB | 6 | 22 | 1.00 | .875 | .500 | 1.00 | .889 |
| Gad1-F-300029 | IDFP | FB,PB | 9 | 19 | 1.00 | .760 | .143 | 1.00 | .769 |
| Gad1-F-300066 | IDFP | FB,PB | 19 | 29 | 1.00 | .585 | .056 | 1.00 | .595 |
| Gad1-F-300099 | IDFP | FB,PB | 3 | 23 | 1.00 | 1.00 | 1.00 | 1.00 | 1.00 |
| Gad1-F-300121 | FB,NO | PB | 12 | 5 | 1.00 | 1.00 | 1.00 | 1.00 | 1.00 |
| Gad1-F-300123 | IDFP | FB,PB | 17 | 35 | 1.00 | 1.00 | 1.00 | 1.00 | 1.00 |
| Gad1-F-300189 | FB,NO | PB | 10 | 12 | 1.00 | 1.00 | 1.00 | 1.00 | 1.00 |
| Gad1-F-300520 | CAL,LH | AL | 11 | 12 | 1.00 | 1.00 | 1.00 | 1.00 | 1.00 |
| Gad1-F-400005 | FB,NO | PB | 11 | 7 | 1.00 | .875 | .909 | 1.00 | .944 |
| Gad1-F-400017 | IDFP | FB,PB | 7 | 26 | 1.00 | 1.00 | 1.00 | 1.00 | 1.00 |
| Gad1-F-400104 | FB,NO | PB | 15 | 2 | 1.00 | .286 | .667 | 1.00 | .706 |
| Gad1-F-400312 | FB,NO | PB | 20 | 6 | 1.00 | 1.00 | 1.00 | 1.00 | 1.00 |
| Gad1-F-400385 | FB,NO | PB | 17 | 11 | 1.00 | 1.00 | 1.00 | 1.00 | 1.00 |
| Gad1-F-400400 | PB | CCP,VMP | 7 | 26 | 0.00 | .731 | 0.00 | .950 | .704 |
| Gad1-F-500035 | IDFP | FB,PB | 13 | 58 | 1.00 | 1.00 | 1.00 | 1.00 | 1.00 |
| Gad1-F-500065 | IDFP | FB,PB | 8 | 35 | 1.00 | .829 | .125 | 1.00 | .833 |
| Gad1-F-500299 | CAL,LH | AL | 13 | 80 | 1.00 | 1.00 | 1.00 | 1.00 | 1.00 |
| Gad1-F-500312 | CAL,LH | AL | 13 | 68 | 1.00 | 1.00 | 1.00 | 1.00 | 1.00 |
| Gad1-F-500661 | CAL,LH | AL | 17 | 11 | 1.00 | 1.00 | 1.00 | 1.00 | 1.00 |
| Gad1-F-600003 | FB,NO | PB | 5 | 6 | .400 | 0.00 | .800 | 0.00 | .364 |
| Gad1-F-600006 | FB,NO | PB | 8 | 4 | 1.00 | 1.00 | 1.00 | 1.00 | 1.00 |
| Gad1-F-600025 | EB,NO | PB | 11 | 9 | 1.00 | 1.00 | 1.00 | 1.00 | 1.00 |
| Gad1-F-600033 | IDFP | FB,PB | 13 | 26 | 1.00 | .774 | .417 | 1.00 | .806 |
| Gad1-F-600077 | FB,NO | PB | 7 | 4 | 1.00 | 1.00 | 1.00 | 1.00 | 1.00 |
| Gad1-F-600081 | IDFP | FB,PB | 10 | 34 | 1.00 | .850 | .400 | 1.00 | .864 |
| Gad1-F-600084 | IDFP,PB | EB | 9 | 24 | 1.00 | .786 | .333 | 1.00 | .806 |
| Gad1-F-700120 | CAL,LH | AL | 35 | 20 | 1.00 | 1.00 | 1.00 | 1.00 | 1.00 |
| Gad1-F-700125 | FB,NO | PB | 10 | 5 | 1.00 | 1.00 | 1.00 | 1.00 | 1.00 |
| Gad1-F-700150 | CAL,LH | AL | 10 | 9 | 1.00 | 1.00 | 1.00 | 1.00 | 1.00 |
| Gad1-F-700275 | CAL,LH | AL | 8 | 24 | 1.00 | 1.00 | 1.00 | 1.00 | 1.00 |
| Gad1-F-800013 | IDFP | FB,PB | 18 | 55 | 1.00 | 1.00 | 1.00 | 1.00 | 1.00 |
| Gad1-F-800025 | IDFP | FB,PB | 12 | 42 | 1.00 | 1.00 | 1.00 | 1.00 | 1.00 |
| Gad1-F-800046 | FB,NO | PB | 20 | 5 | 1.00 | 1.00 | 1.00 | 1.00 | 1.00 |
| Gad1-F-800113 | FB,NO | PB | 17 | 12 | 1.00 | 1.00 | 1.00 | 1.00 | 1.00 |
| Gad1-F-800139 | FB,NO | PB | 19 | 5 | 1.00 | 1.00 | 1.00 | 1.00 | 1.00 |
| Gad1-F-900011 | IDFP | FB,PB | 13 | 37 | 1.00 | .949 | .833 | 1.00 | .959 |
| Gad1-F-900035 | CAL,LH | AL | 11 | 9 | 1.00 | 1.00 | 1.00 | 1.00 | 1.00 |
| npf-F-100003 | CAL,LH | AL | 12 | 21 | 1.00 | 1.00 | 1.00 | 1.00 | 1.00 |
| npf-F-100004 | CAL,LH | AL | 14 | 34 | 1.00 | 1.00 | 1.00 | 1.00 | 1.00 |
| npf-F-100009 | CAL,LH | AL | 12 | 18 | 1.00 | 1.00 | 1.00 | 1.00 | 1.00 |
| npf-F-100010 | CAL,LH | AL | 16 | 15 | 1.00 | 1.00 | 1.00 | 1.00 | 1.00 |
| npf-F-100011 | CAL,LH | AL | 17 | 21 | 1.00 | 1.00 | 1.00 | 1.00 | 1.00 |
| npf-F-200001 | CAL,LH | AL | 13 | 11 | 1.00 | 1.00 | 1.00 | 1.00 | 1.00 |
| npf-F-200003 | CAL,LH | AL | 20 | 17 | 1.00 | 1.00 | 1.00 | 1.00 | 1.00 |
| npf-F-200008 | CAL,LH | AL | 13 | 16 | 1.00 | 1.00 | 1.00 | 1.00 | 1.00 |
| npf-F-200018 | CAL,LH | AL | 17 | 28 | 1.00 | 1.00 | 1.00 | 1.00 | 1.00 |
| npf-F-200042 | CAL,LH | AL | 12 | 16 | 1.00 | 1.00 | 1.00 | 1.00 | 1.00 |
| npf-F-200044 | CAL,LH | AL | 15 | 11 | 1.00 | 1.00 | 1.00 | 1.00 | 1.00 |
| npf-M-100010 | CAL,LH | AL | 13 | 18 | 1.00 | 1.00 | 1.00 | 1.00 | 1.00 |
| Tdc2-F-200009 | IDFP | FB,PB | 4 | 25 | 1.00 | 1.00 | 1.00 | 1.00 | 1.00 |
| Tdc2-F-300003 | IDFP | FB,PB | 8 | 29 | 1.00 | 1.00 | 1.00 | 1.00 | 1.00 |
| Tdc2-F-300014 | IDFP | FB,PB | 5 | 34 | 1.00 | 1.00 | 1.00 | 1.00 | 1.00 |
| Tdc2-F-300036 | IDFP | FB,PB | 5 | 30 | 1.00 | 1.00 | 1.00 | 1.00 | 1.00 |
| Tdc2-F-300042 | IDFP | FB,PB | 3 | 34 | 1.00 | 1.00 | 1.00 | 1.00 | 1.00 |
| Tdc2-F-400002 | IDFP | FB,PB | 3 | 22 | 1.00 | 1.00 | 1.00 | 1.00 | 1.00 |
| Tdc2-F-400009 | IDFP | FB,PB | 2 | 32 | 1.00 | 1.00 | 1.00 | 1.00 | 1.00 |
| Tdc2-F-400026 | CAL,LH | AL,VMP | 157 | 269 | 1.00 | 1.00 | 1.00 | 1.00 | 1.00 |
| Tdc2-F-500000 | IDFP | FB,PB | 6 | 32 | 1.00 | 1.00 | 1.00 | 1.00 | 1.00 |
| Tdc2-F-600000 | IDFP | FB,PB | 4 | 34 | 1.00 | 1.00 | 1.00 | 1.00 | 1.00 |
| VGlut-F-000485 | CAL,LH | AL | 5 | 81 | 1.00 | 1.00 | 1.00 | 1.00 | 1.00 |
| VGlut-F-200566 | CAL,LH | AL | 27 | 62 | 1.00 | 1.00 | 1.00 | 1.00 | 1.00 |
| VGlut-F-200574 | CAL,LH | AL | 9 | 24 | 1.00 | 1.00 | 1.00 | 1.00 | 1.00 |
| VGlut-F-300243 | FB,NO | PB | 9 | 3 | 1.00 | 1.00 | 1.00 | 1.00 | 1.00 |
| VGlut-F-300517 | FB,NO | PB | 18 | 7 | 1.00 | .353 | .353 | 1.00 | .522 |
| VGlut-F-300596 | CAL,LH | AL | 18 | 10 | 1.00 | 1.00 | 1.00 | 1.00 | 1.00 |
| VGlut-F-400664 | CAL,LH | AL | 13 | 9 | 1.00 | 1.00 | 1.00 | 1.00 | 1.00 |
| VGlut-F-500626 | CAL,LH | AL | 9 | 7 | 1.00 | 1.00 | 1.00 | 1.00 | 1.00 |
| VGlut-F-600146 | CAL,LH | AL | 8 | 18 | 1.00 | 1.00 | 1.00 | 1.00 | 1.00 |
| VGlut-F-600243 | CAL,LH | AL | 14 | 5 | 1.00 | 1.00 | 1.00 | 1.00 | 1.00 |
| VGlut-F-600248 | CAL,LH | AL | 13 | 19 | 1.00 | 1.00 | 1.00 | 1.00 | 1.00 |
| VGlut-F-600751 | CAL,LH | AL | 6 | 13 | 1.00 | 1.00 | 1.00 | 1.00 | 1.00 |
| VGlut-F-700285 | CAL,LH | AL | 7 | 15 | 1.00 | 1.00 | 1.00 | 1.00 | 1.00 |
| VGlut-F-700494 | CAL,LH | AL | 14 | 8 | 1.00 | 1.00 | 1.00 | 1.00 | 1.00 |
| VGlut-F-700547 | CAL,LH | AL | 16 | 28 | 1.00 | 1.00 | 1.00 | 1.00 | 1.00 |
| VGlut-F-800097 | CAL,LH | AL | 7 | 10 | 1.00 | 1.00 | 1.00 | 1.00 | 1.00 |